\documentclass[11pt]{article}
\usepackage{amsmath}
\usepackage{graphicx}
\usepackage{amssymb}
\usepackage{amsfonts}
\usepackage{latexsym}
\usepackage{color}
\usepackage{verbatim,cite}

\renewcommand{\vec}[1]{{\bf #1}}

\def\beq{\begin{eqnarray}}
\def\eeq{\end{eqnarray}}
\def\ln{\,\mbox{ln}\,}

\def\si{\sigma}

\def\Om{\Omega}

\newcommand{\CC}{\Lambda}
\newcommand{\rL}{\rho_{\CC}}
\newcommand{\rLo}{\rho_{\CC}^0}
\newcommand{\drL}{\delta\rho_{\CC}}

\newcommand{\OM}{\Omega_m}
\newcommand{\OR}{\Omega_r}
\newcommand{\ORo}{\Omega_r^0}
\newcommand{\omm}{\omega_m}
\newcommand{\Omo}{\Omega_m^0}
\newcommand{\OL}{\Omega_{\Lambda}}
\newcommand{\OLo}{\Omega_{\Lambda}^0}

\newcommand{\rco}{\rho_{c}^0}
\newcommand{\rmo}{\rho_{m}^0}

\newcommand{\rM}{\rho_m}
\newcommand{\pM}{p_m}

\newcommand{\mincir}{\raise
-3.truept\hbox{\rlap{\hbox{$\sim$}}\raise4.truept\hbox{$<$}\ }}
\newcommand{\magcir}{\raise
-3.truept\hbox{\rlap{\hbox{$\sim$}}\raise4.truept\hbox{$>$}\ }}
\newcommand{\newtext}[1]{\textcolor{black}{#1}}

\begin{document}

\begin{center}
{\Large\sc Cosmic perturbations with running $G$ and $\Lambda$}
\vskip 6mm

{\bf Javier Grande, Joan Sol\`a}\\

HEP Group, Dept. ECM and Institut de Ci{\`e}ncies del Cosmos\\
Univ. de Barcelona, Av. Diagonal 647, E-08028 Barcelona, Catalonia,
Spain

\vspace{0.2cm}

\quad {\bf Julio C. Fabris} \\

Departamento de F{\'\i}sica -- CCE,
\\
Univ. Federal do Espir{\'\i}to Santo,  CEP 29060-900,
Vit\'oria, ES, Brazil\\

\vspace{0.2cm}

 \quad {\bf Ilya L. Shapiro}

Departamento de F\'{\i}sica - ICC
\\
Univ. Federal de Juiz de Fora,  CEP 36036-330, MG, Brazil\\

\vspace{0.5cm}

E-mail: jgrande@ecm.ub.es, sola@ecm.ub.es, fabris@pq.cnpq.br,
shapiro@fisica.ufjf.br

\vskip 2mm
\end{center}

\vskip 6mm


\begin{quotation}
\noindent {\large\it Abstract.} Cosmologies with running
cosmological term $\rL$ and gravitational Newton's coupling $G$ may
naturally be expected if the evolution of the universe can
ultimately be derived from the first principles of Quantum Field
Theory or String Theory. For example, if matter is conserved and the
vacuum energy density varies quadratically with the expansion rate
as $\rL(H)=n_0+n_2\,H^2$, with $n_0\neq 0$  (a possibility that has
been advocated in the literature within the QFT framework), it can
be shown that G must vary logarithmically (hence very slowly) with
$H$. {In this paper, we derive the general cosmological perturbation
equations for models with variable $G$ and $\rL$ in which the
fluctuations $\delta G$ and $\delta\rL$ are explicitly included. We
demonstrate that, if matter is covariantly conserved, the late
growth of matter density perturbations is independent of the
wavenumber $k$. Furthermore, if $\rL$ is negligible at high
redshifts and $G$ varies slowly, we find that these cosmologies
produce a matter power spectrum with the same shape as that of the
$\CC$CDM model, thus predicting the same basic features on structure
formation. Despite this shape indistinguishability, the free
parameters of the variable $G$ and $\rL$ models can still be
effectively constrained from the observational bounds on the
spectrum amplitude.}

 \vskip 3mm

PACS: $\,$ 95.36.+x $\,\,$  04.62.+v $\,\,$ 11.10.Hi \vskip 3mm

\end{quotation}

\vskip 12mm

\newpage
\section{\large\bf Introduction}

The current ``standard model'' (or ``concordance model'') of our
universe, being an homogeneous and isotropic FLRW cosmological
model, consists of a remarkably small number of ingredients, to wit:
matter, radiation and a cosmological constant (CC) term, $\CC$. The
first two ingredients are dynamical and vary (decrease fast) with
the cosmic time $t$ whereas the third, $\CC$, remains strictly
constant. After many years of theoretical insight (cf. e.g. the
reviews\,\cite{Weinberg89,Peebles03} and references therein), the
situation of the original FLRW cosmological models has not changed
much, in the sense that we have not been able to make any
fundamental advance in the comprehension of the relationship between
the matter energy density and the CC. Still, we have performed a
major accomplishment at the phenomenological level by simultaneously
fitting the modern independent data sets emerging from LSS galaxy
surveys, supernova luminosities and the cosmic microwave background
(CMB) anisotropies\,
\cite{Cole:2005sx,Teg04,Reid:2009xm,Riess:2004nr,Spergel07}. On the
basis of this successful fit, in which $\CC$ enters as a free
parameter, one claims that a non-vanishing and positive cosmological
constant has been measured. Nevertheless, we still don't know what
is the true meaning of the fitted parameter $\CC$. Assuming that
Newton's gravitational coupling $G$ is strictly constant, the
combined set of observational data determine the value
\begin{equation}\label{CCvalue}
\rL=\frac{\CC}{8\pi G}\simeq (2.3\times 10^{-3}\,eV)^4\,,
\end{equation}
which we interpret as the vacuum energy density. It corresponds to
$\OLo=\rL/\rco\simeq 0.7$ when the $\rL$ density is normalized with
respect to the current critical density $\rco=(3\sqrt{h}\,\times
10^{-3}\, eV)^4$ -- for a reduced Hubble constant value of $h\simeq
0.70$. Assuming (in the light of the same observational data) that
the universe is spatially flat, this means that the current matter
density normalized to the critical density is $\Omo=\rmo/\rco\simeq
0.3$.

All {the efforts} to deduce the value of the energy density
(\ref{CCvalue}) -- a very small one for all particle physics
standards, except if a very light neutrino mass is the only particle
involved\,\cite{ShapSol00} -- have failed up to now. The main
stumbling block to a solution is {the fact that any} approach based
on the fundamental principles of quantum field theory (QFT) or
string theory lead to some explicit or implicit form of severe
fine-tuning among the parameters of the theory. The reason for this
is that these theoretical descriptions are plagued by large
hierarchies of energy scales associated to the existence of many
possible vacua.

This difficulty became clear already from the first attempts to
treat the dark energy component as a dynamical scalar field
\cite{OldScalar}. The idea was to let {such a} field select
automatically the vacuum state in a dynamical way, especially one
with zero value of the energy density. More recently, this approach
took the popular form of a ``quintessence'' field {slowly} rolling
down its potential and has adopted many different
faces\,\cite{Quintessence} -- for a review, see \cite{Peebles03}.
But the situation at present is even harder because the quintessence
field -- or, {more generally}, a dark energy (DE) field -- should be
able to naturally choose {not zero but} the very small number
(\ref{CCvalue}) as its ground state.

Despite {the fact} that a working dynamical mechanism able to choose
the correct vacuum state has to be found yet, another important
motivation for the quintessence models is that they aim at
explaining the puzzling coincidence between the present value of
$\rL$ and the value of the matter density, $\rmo$. In other words,
why $\OLo/\Omo={\cal O}(1)$? Arguably, a dynamical DE should be a
starting point to understand this puzzle. Detailed analyses of these
models exist in the literature, including their confrontation with
the data\,\cite{Jassal,SR,Xin,SVJ}.

On the other hand, the possibility that the cosmological term is a
running quantity, which could be sensitive to the quantum matter
effects, seems a more appealing ansatz, as it could provide an
interface between QFT and cosmology\,\cite{ShapSol00,ShapSol02}.
This fundamental possibility has been recently emphasized in
\cite{ShapSol09a}{-- for a review, see \cite{ShapSolRev07}.}
Actually, the analysis of the various observational data show (see
Ref.\cite{BPS09,CCfit}) that a wide class of dynamical CC models
models are indeed able to fit the combined observations to a level
of accuracy comparable to the standard $\CC$CDM model. In some
cases, the dynamical nature of $\CC$ allows these models to provide
a clue to the coincidence problem\,\cite{LXCDM12,LXCDM08}, and maybe
eventually to the full cosmological constant
problem\,\cite{FSS09abc}.

In general, in this kind of dynamical CC scenarios we have
$\CC=\CC(\xi)$ where $\xi=\xi(t)$ is a cosmological variable that
evolves with time, and therefore ultimately $\CC=\CC(t)$ is also
time varying. Originally, these models were purely phenomenological,
with no relation to QFT \,\cite{oldvarCC1}. {Nowadays we know that
not all models of this kind are allowed, and the fact that the
observational data can discriminate which of them are good
candidates and which are not so good gives some insight into} the
function $\xi=\xi(t)$\,\cite{BPS09}. A particularly interesting
class of variable CC models are those in which the gravitational
coupling $G$ changes very slowly (logarithmically) with the
expansion of the universe\,\cite{SSS04,Fossil07}. This scenario is
possible e.g. if matter is covariantly conserved.

In practice, $\xi$ could be the expansion rate $H=H(t)$, the scale
factor $a=a(t)$, the matter energy density, etc. In this paper, we
shall assume that $\xi=H$, similarly as it was done for models with
a variable $\CC$ {interacting} with
matter\,\cite{ShapSol02,Babic,SS12}. With this ansatz, we shall
study the impact on the structure formation, showing that these
models predict the same shape for the matter power spectrum as the
$\CC$CDM model, a fact that would not apply e.g. for variable CC
models in which the CC decays into matter. We find this feature
remarkable and we shall explore its possible consequences and also
some possible phenomenological tests of this kind of models.

The paper is organized as follows. In the next section, we classify
the possible scenarios with variable $\rL$ and $G$. In section 3, we
present the general set of coupled perturbation equations involving
$\delta\rL$ and $\delta G$, showing that the late growth of matter
fluctuations becomes independent of the scale $k$. A class of models
with variable cosmological term as a series function of the Hubble
rate is introduced in section 4. In section 5, we concentrate on a
particular model in this class, which is well motivated from the QFT
point of view, and analyze the constraints imposed by primordial
nucleosynthesis and structure formation. In the last section, we
present and discuss our conclusions. Finally, an appendix is
included at the end where we discuss gauge issues.

\section{Generic models with variable cosmological parameters}\label{sec:GenericModels}

\qquad We start from Einstein's equations in the presence of the
cosmological constant term,
\begin{equation}
R_{\mu \nu }-\frac{1}{2}g_{\mu \nu }R= 8\pi
G\,(T_{\mu\nu}+g_{\mu\nu}\,\rL)\,, \label{EE}
\end{equation}
where $T_{\mu\nu}$ is the ordinary energy-momentum tensor associated
to isotropic matter and radiation, and $\rL$ represents the vacuum
energy density associated to the CC.  Let us now contemplate the
possibility that $G=G(t)$ and $\CC=\CC(t)$ can be both functions of
the cosmic time within the context of the FLRW  cosmology. It should
be clear that the very precise measurements of $G$ existing in the
literature refer only to distances within the solar system and
astrophysical systems. In cosmology these scales are immersed into
much larger scales (galaxies and clusters of galaxies) which are
treated as point-like (and referred to as ``fundamental observers'',
comoving with the cosmic fluid). Therefore, the variations of $G$ at
the cosmological level could only be observed at much larger
distances where at the moment we have never had the possibility to
make direct experiments. In practice, the potential variation of
$G=G(t)$ and $\CC=\CC(t)$ should be expressed in terms of a possible
cosmological redshift dependence of these functions, $G=G(z)$ and
$\CC=\CC(z)$.

Let us consider the various possible scenarios for variable
cosmological parameters that appear when we solve Einstein's
equations (\ref{EE}) in the flat FLRW metric,
$ds^{2}=dt^{2}-a^{2}(t)d\vec{x}^{2}$. To start with, one finds
Friedmann's equation with non-vanishing $\rL$, which provides
Hubble's expansion rate $H=\dot{a}/a$ ($\dot{a}\equiv da/dt$) as a
function of the matter and vacuum energy densities:
\begin{equation}\label{Friedmann}
H^2=\frac{8\pi G}{3}(\rM+\rL)\,.
\end{equation}
On the other hand, the general Bianchi identity of the Einstein
tensor in (\ref{EE}) leads to
\begin{equation}\label{GBI}
\bigtriangledown^{\mu}\,\left[G\,(T_{\mu\nu}+g_{\mu\nu}\,\rL)\right]=0\,.
\end{equation}
Using the FLRW metric explicitly, the last equation results into the
following ``mixed'' local conservation law:
\begin{equation}\label{BianchiGeneral}
\frac{d}{dt}\,\left[G(\rM+\rL)\right]+3\,G\,H\,(\rM+\pM)=0\,.
\end{equation}
If {$\dot{\rho}_{\CC}\neq 0$}, then $\rM$ is not generally conserved
as there may be {transfer} of energy from matter-radiation into the
variable $\rL$ or vice versa (including a possible contribution from
a variable $G$, if $\dot{G}\neq 0$). Thus this law mixes, in
general, the matter-radiation energy density with the vacuum energy
$\rL$. However, the following particular scenarios are possible:
\begin{itemize}

\item i) $G=$const. {and} $\rL=$const. This is the standard case of
$\CC$CDM cosmology, implying the local covariant conservation law of
matter-radiation:
\begin{equation}\label{standardconserv}
\dot{\rho}_m+3\,H\,(\rM+\pM)=0;
\end{equation}

\item  ii) $G=$const
{and} $\dot{\rho}_{\CC}\neq 0$, in which case
Eq.(\ref{BianchiGeneral}) leads to the mixed conservation law
\begin{equation}\label{mixed conslaw}
\dot{\rho}_{\CC}+\dot{\rho}_m+3\,H\,(\rM+\pM)=0;
\end{equation}

\item iii) $\dot{G}\neq 0$ {and} $\rL=$const, implying
$\dot{G}(\rM+\rL)+G[\dot{\rho}_m+3H(\rM+\pM)]=0$;

\item iv) $\dot{G}\neq 0$ {and} $\dot{\rho}_{\CC}\neq 0$. In this case, if
we assume the standard local covariant conservation of
matter-radiation, i.e Eq.\,(\ref{standardconserv}), it is easy to
see that Eq.\,(\ref{BianchiGeneral}) boils down to
\begin{equation}\label{Bianchi}
(\rM+\rL)\dot{G}+G\dot{\rL}=0\,.
\end{equation}

\end{itemize}
Notice that only in cases i) and iv) matter is covariantly
self-conserved, meaning that matter evolves according to the
{solution of Eq.\,(\ref{standardconserv}):}
\begin{equation}\label{solstandardconserv}
\rM(a)=\rM^0\,a^{-\alpha_m}=\rM^0\,(1+z)^{\alpha_m}\,,\ \ \
\alpha_m=3(1+\omm)\,,
\end{equation}
where $\rM^0$ is the current matter density and $\omm=\pM/\rM=0,1/3$
are the equation of state (EOS) parameters for cold and relativistic
matter, respectively. We have expressed the result
(\ref{solstandardconserv}) in terms of the scale factor $a=a(t)$ and
the cosmological redshift $z=(1-a)/a$.

In cases ii) and iii), instead, matter is not conserved (if one of
the two parameters $\rL$ or $G$ indeed is to be variable,
respectively). Explicit cosmological models with variable parameters
as in case ii) have been investigated in detail
in\,\cite{ShapSol02,CCfit,SS12}. Cosmic perturbations of this model
have been considered in
\cite{Fabris:2006gt,Grande:2007wj,vdecay,WangMeng05,Toribio09}. Case
iii) has been studied at the background level in
\cite{Guberina:2006fy}. Finally, the background evolution of case
iv) has been studied in different contexts in
Ref.\cite{SSS04,Fossil07}\,\footnote{Some potential astrophysical
implications of scenario iv) have been addressed first in
\cite{SSS04} and recently in \cite{Davi09}.}. In the next section,
we shall address the calculation of cosmic perturbations in a
general model where $\rL$ and $G$ can evolve with the expansion, and
we will then specialize the set of equations for the type iv) model
in which matter is covariantly conserved. Only in section
\ref{sec:QFTvac} we will further narrow down the obtained set of
perturbation equations for a concrete model with running
cosmological parameters\,\cite{SSS04}.

\qquad

\section{ Perturbations with variable $\CC$ and
$G$}\label{sec:general_perturb}

\qquad For the analysis of the cosmic perturbations in the general
running $\rL$ and $G$ model iv) of the previous section, we have to
perturb all parts of Einstein's equations that may evolve with time;
namely the metric, the energy-momentum tensor for both matter and
vacuum, and finally we must perturb also the gravitational constant.
Einstein's equations (\ref{EE}) can be conveniently cast as follows:
\begin{equation}\label{Einsteins}
R_{\mu\nu}=8\pi
G\left(T_{\mu\nu}-\frac{1}{2}g_{\mu\nu}T^{\lambda}_{\lambda}\right)\,.
\end{equation}
As a background metric, we use the flat FLRW metric:
\begin{equation}
ds^2=g_{\mu\nu}dx^{\mu}dx^{\nu}=dt^2-a^2(t)\,\delta_{ij}dx^idx^j\,.
\end{equation}
In perturbing it, $ g_{\mu\nu}\rightarrow g_{\mu\nu}+ h_{\mu\nu}$,
we adopt here the synchronous gauge\,\footnote{See the Appendix for
an alternative calculation in the
\newtext{Newtonian} gauge.} ($h_{00}=h_{0i}=0$). The total energy
momentum tensor of the cosmological fluid can be written as the sum
of the matter and vacuum contributions:
\begin{equation}
T^{\mu}_{\nu}= T^{\mu}_{\nu\textrm{ (matter)}}+T^{\mu}_{\nu\textrm{
($\Lambda$)}}= -p_T\delta^{\mu}_{\nu}+(\rho_{\small T}+p_{\small
T})U^{\mu}U_{\nu}\,, \label{enemom}
\end{equation}
where
\begin{eqnarray}
\rho_{\small
T}&=&\rho_m+\rho_{\Lambda} \nonumber \\
p_{\small T} &=& p_m + p_{\Lambda}=\omm\rho_m - \rho_{\Lambda}\,,
\label{ptot}
\end{eqnarray}
(notice that $\omega_{\CC}\equiv p_{\CC}/\rho_{\CC}=-1$ even if the
CC is running). As for the density and pressure, we introduce the
perturbations in the usual form, except that we have to include also
the perturbation for the vacuum energy density because, in the
present context, $\CC$ is an evolving variable. Thus, we have
$\rho_i\rightarrow \rho_i+\delta\rho_i\,,\ p_i\rightarrow p_i+\delta
p_i$, where $i=m,\Lambda$. As warned above, if we allow for a
variable gravitational coupling $G$, we must also consider
perturbations for it:
\begin{equation}\label{perturbG}
G\rightarrow G+\delta G\,.
\end{equation}
Obviously, the perturbations $\delta G$ and $\drL$ are the two most
distinguished dynamical components of the present cosmic
perturbation analysis, {as they are both absent in the $\CC$CDM
model.}

{Finally, we} have to perturb the 4-velocity of the matter
particles, $U^{\mu}\rightarrow U^{\mu}+\delta U^{\mu}$. For an
observer moving with the fluid (i.e. a comoving observer) it reads:
$U^{\mu}=(1,\delta U^i)$.

We are now ready to find the perturbed equations of motion for this
running model. As the fundamental equations describing the
perturbations for our model, we will take: i) the 00 component of
the Einstein equations (\ref{Einsteins}); and ii) the generalized
Bianchi identity (\ref{GBI}), which splits into energy conservation
(notice that $\dot{G}\neq 0$ in the present framework)
\begin{equation}\label{conservEnergy}
\nabla_{\mu}\left(GT^{\mu}_0\right)=\partial_{\mu}\left(GT^{\mu}_0\right)+
G\left[\Gamma^{\mu}_{\sigma\mu}T^{\sigma}_0-\Gamma^{\sigma}_{\mu
0}T^{\mu}_{\sigma}\right]=0\,,
\end{equation}
and momentum conservation:
\begin{equation}\label{conservMomentum}
\nabla_{\mu}\left(GT^{\mu}_i\right)=
\partial_{\mu}\left(GT^{\mu}_i\right)+ G\left[\Gamma^{\mu}_{\sigma\mu}T^{\sigma}_i-\Gamma^{\sigma}_{\mu
i}T^{\mu}_{\sigma}\right]=0\,.
\end{equation}
As for the 00 component of the Einstein equations, and taking into
account that $G$ gets perturbed as in (\ref{perturbG}), we have:
\begin{eqnarray}
\emph{0th order:}\; && -3\frac{\ddot{a}}{a}=4\pi G \left(\rho_T+3 p_T\right)\,,\nonumber\\
\\[-0.1cm]
\emph{1st order:}\; && \dot{\hat{h}}+2H\hat{h}=8\pi
\left[G\left(\delta\rho_T+3\delta p_T\right) + \delta G
\left(\rho_T+3p_T\right)\right]\,, \nonumber
\end{eqnarray}
where we have defined $\displaystyle
\hat{h}=\frac{\partial}{\partial t}\left(\frac{h_{kk}}{a^2}\right)$
and repeated Latin indices are understood to be summed over the
values 1, 2, 3. As energy-density components, we have matter and the
running cosmological constant, with EOS parameters indicated in
(\ref{ptot}), and thus the perturbed equation takes on the form:
\begin{equation}\label{generalEq1}
\dot{\hat{h}}+2H\hat{h}=8\pi\Big\{G\big[(1+3\omega_m)
\delta\rho_m+\delta\rho_{\Lambda}+3\delta p_{\Lambda}\big]+\delta
G\big[(1+3\omega_m)\rho_m-2\rho_{\Lambda}\big]\Big\}\,.
\end{equation}
Let us now work out the equation of local covariant conservation of
the energy, Eq.(\ref{conservEnergy}). After a straightforward
calculation, the final equations read as follows. On the one hand,
\begin{eqnarray}\label{conservEnergy0}
\emph{0th
order:}\;&&\dot{G}(\rho_m+\rho_{\Lambda})+G(\dot{\rho}_m+\dot{\rho}_{\Lambda})+3GH\rho_m(1+\omega_m)=0\,.
\end{eqnarray}
If we assume local covariant conservation of matter,
$\nabla_{\mu}T^{\mu}_{0\textrm{ (matter)}}=0$, i.e.
Eq.\,(\ref{standardconserv}), we see that (\ref{conservEnergy0})
indeed reduces to Eq.\,(\ref{Bianchi}), as it should be. On the
other hand,
\begin{eqnarray}\label{conservEnergy1}
\emph{1st
order:}\;&&G\left[\delta\dot{\rho}_m+\delta\dot{\rho}_{\Lambda}+\rho_m(1+\omega_m)\left(\theta-\frac{\hat{h}}{2}\right)+
3H\big[(1+\omega_m)\delta\rho_m+(\delta\rho_{\Lambda}+\delta p_{\Lambda})\big]\right]+\nonumber\\
&& \quad +\dot{G}(\delta\rho_m+\delta\rho_{\Lambda})
+\big[\dot{\rho}_m+\dot{\rho}_{\Lambda} +
3H(1+\omega_m)\rho_m\big]\delta G +  (\rho_m+\rho_{\Lambda})\delta
\dot{G}=0\,.
\end{eqnarray}
In the previous equation, we have introduced the variable
\begin{equation}
\theta\equiv\nabla_{\mu}\delta U^{\mu}=\partial_i\delta U^i\,,
\end{equation}
which represents the perturbation on the matter velocity gradient,
and we have used $\delta U^0=0$.
If we invoke once more the local covariant conservation of matter,
both at 0th order -- see Eq.\,(\ref{standardconserv}) -- and at 1st
order,
\begin{equation}\label{generalEq2}
\delta\dot{\rho}_m+\rho_m(1+\omega_m)\left(\theta-\frac{\hat{h}}{2}\right)+3H(1+\omega_m)\delta\rho_m=0\,,
\end{equation}
equation (\ref{conservEnergy1}) can be further reduced to the
simpler form:
\begin{equation}\label{generalEq3}
G\big[\delta\dot{\rho}_{\Lambda}+3H\big(\delta\rho_{\Lambda}+\delta
p_{\Lambda}\big)\big]+\dot{G}(\delta\rho_m+\delta\rho_{\Lambda})+\dot{\rho}_{\Lambda}
\delta G + (\rho_m+\rho_{\Lambda})\delta\dot{G}=0\,.
\end{equation}
Finally, let us work out the equation of local covariant
conservation of momentum, Eq.(\ref{conservMomentum}). In this
instance, there are no 0th order terms (in the background, the
momentum conservation is trivial). The perturbed result reads
\begin{eqnarray}
a^2\partial_t[G \rho(1+\omega)\delta U^i] + 5a\dot{a}G
\rho(1+\omega)\delta U^i + G\partial_i(\delta p) +
p\partial_i(\delta G)=0\,,
\end{eqnarray}
{where it is understood that we should sum over all the energy
components, in this case, matter and $\CC$}. By applying the
operator $\partial_i$ and transforming to the Fourier space, we get:
\begin{eqnarray}
(1+\omega)\left[\dot{G}\rho\theta+G(\dot{\rho}\theta +
\rho\dot{\theta}+5H\rho\theta)\right]- \frac{k^2}{a^2}\left[G\delta
p+\omega\rho\delta G\right]&=&0\,,
\end{eqnarray}
{which, after summing over matter and CC, becomes:}
\begin{equation}\label{perturbconservMomentum}
(1+\omega_m)\left[\dot{G}\rho_m\theta+G(\dot{\rho}_m\theta +
\rho_m\dot{\theta}+5H\rho_m\theta)\right]=
\frac{k^2}{a^2}\left[G(\delta p_{\Lambda}+\omega_m\delta
\rho_m)+(\omega_m\rho_m-\rho_{\Lambda})\delta G\right]\,.
\end{equation}
The part corresponding to matter conservation
$\left(\nabla_{\mu}T^{\mu}_{i\textrm{ (matter)}}=0\right)$ is
\begin{equation}\label{generalEq4}
(1+\omega_m)\left[\dot{\rho}_m\theta +
\rho_m\dot{\theta}+5H\rho_m\theta\right]= \frac{k^2}{a^2}\omega_m
\delta\rho_m\,,
\end{equation}
so that Eq.\,(\ref{perturbconservMomentum}) simplifies as follows:
\begin{equation}\label{generalEq5}
(1+\omega_m)\dot{G}\rho_m\theta=\frac{k^2}{a^2}\left[G\delta
p_{\Lambda}+(\omega_m\rho_m-\rho_{\Lambda})\delta G\right]\,.
\end{equation}
Summarizing, our final set of equations is given by
(\ref{generalEq1}),(\ref{generalEq2}),(\ref{generalEq3}),(\ref{generalEq4})
and (\ref{generalEq5}). It is particularly relevant for our purposes
to rewrite these equations in the matter dominated epoch
($\omega_m=0$) and assuming adiabatic perturbations for the CC
($\delta p_{\Lambda}=-\delta\rho_{\Lambda}$). In a nutshell, we
find:
\begin{eqnarray}
&&\dot{\hat{h}}+2H\hat{h}=8\pi\big[\rho_m-2\rho_{\Lambda}\big]\delta G+ 8\pi G \big[\delta\rho_m-2\delta\rho_{\Lambda}\big]\label{fin1}\\
&& \delta\dot{\rho}_m+\rho_m\left(\theta-\frac{\hat{h}}{2}\right)+3H \delta\rho_m=0\label{fin2}\\
&&\dot{\theta}+2H\theta=0\label{fin3}\\
&& \delta\dot{G}(\rho_m+\rho_{\Lambda})+\delta G\dot{\rho}_{\Lambda}
+
\dot{G}(\delta\rho_m+\delta\rho_{\Lambda})+ G\delta\dot{\rho}_{\Lambda}=0\label{fin4}\\
&& k^2\left[G\delta\rho_{\Lambda}+\rho_{\Lambda}\delta
G\right]+a^2\rho_m\dot{G}\theta=0\,.\label{fin5}
\end{eqnarray}
Since Eq.\,(\ref{fin3}) will be important for the subsequent
considerations, let us note that it follows from
Eq.\,(\ref{generalEq4}) upon using the matter conservation equation
(\ref{standardconserv}).

We are now ready to derive an important result that applies to all
cosmological models of the type iv) in section
\ref{sec:GenericModels}, i.e. models with variable $\CC$ and $G$ in
which matter is covariantly conserved, to wit: for all these models,
the perturbation equations that we have just derived do \emph{not}
depend in fact on the wavenumber $k$. To prove this remarkable
result is very easy at this stage of the discussion, as it ensues
immediately from equations\,(\ref{fin3}) and (\ref{fin5}). Indeed,
if we change the variable from cosmic time $t$ to the scale factor
$a$ (which is readily done by using $\displaystyle \dot{f}\equiv
df/dt=aH df/da\equiv aHf'$), equation (\ref{fin3}) can be cast as
\begin{equation}
\theta'+\frac{2}{a}\theta=0\,.
\end{equation}
Therefore,
\begin{equation}\label{theta2}
\theta=\theta_0\, a^{-2}\,.
\end{equation}
It follows that the second term on the \textit{l.h.s.} of
Eq.\,(\ref{fin5}) behaves as
$a^2\rho_m\dot{G}\theta=\theta_0\,\rho_m\dot{G}$. Notice that
$\theta_0$ is the perturbation of the matter velocity at present,
which is of course much smaller than any value $\theta_i$ that this
variable can take early on, more specifically at any time after the
transfer function regime has finished (see below). Obviously,
$\theta_i=\theta_0\, a_i^{-2}\ll \theta_0$ (where $a_i\ll a_0=1$).
Thus, taking into account that the matter perturbations
$\delta\rho_m$ are growing rather than decaying, $\theta$ will be
comparatively negligible. Let us also note that the $\theta$-term in
Eq.\,(\ref{fin5}) is multiplied by the time derivative of $G$, which
in all reasonable models (in particular, the one we explore in
section \ref{sec:numerical}) should be small. Finally, if we divide
Eq.\,(\ref{fin5}) by $k^2$ and take into account that in practice we
are interested in
\newtext{deep sub-horizon scales}, i.e. scales $\lambda$ that satisfy
$\lambda\,a\ll H^{-1}$ (or, equivalently, $k\gg Ha$), it follows
that the the $\theta$-term  in Eq.\,(\ref{fin5}) is entirely
negligible. Thus, in all practical respects, this is tantamount to
set $\theta(a)=0\ (\forall a)$. The outcome is that the full set of
perturbation equations becomes independent of $k$, as
announced\,\footnote{This result is similar to the model of
Ref.\,\cite{LXCDM08}, where matter is also covariantly conserved.}.
To better assess the physical significance of this important result,
we have repeated the calculation in another gauge (the
\newtext{Newtonian} or longitudinal gauge) and we have obtained the same
result for scales well below the horizon (see the Appendix at the
very end for a summarized presentation and discussion).

Having shown that the shape of the spectrum is not distorted at late
times in our model, we may ask ourselves if, in contrast, there are
additional sources of wavenumber dependence at early times. Recall
that the transfer function $T(k)$ parameterizes the important
dynamical effects that the various $k$-modes undergo at the early
epochs\,\cite{Dodelson}. More specifically, it accounts for the
non-trivial evolution of the primordial perturbations through the
epochs of horizon crossing and radiation/matter transition (the
latter occurs at the so-called ``equality time'', $t_{\rm eq}$). A
most important feature in the $\CC$CDM case is that the scale
dependence is fully encoded in $T(k)$, and so the spectrum of the
standard model evolves without any further distortion during most of
the matter epoch till the present time.

We may ask ourselves if the evolution of our model with running
$\CC$ and $G$ at the early epochs follows also the same pattern as
the $\CC$CDM, or if its dynamics could imprint some significant
modification on the structure of $T(k)$. In such a case, a
distortion of the power spectrum with respect to the $\CC$CDM would
be generated at early times and it would freely propagate (i.e.
without further modifications) till the present days and act as a
kind of signature of the model. However, we deem that this
possibility is not likely to be the case, for in the kind of models
under consideration there is no production/decay of matter or
radiation. Thus, the value of the scale factor at equality, $a_{\rm
eq}=a(t_{\rm eq})$, must be identical to that of the standard
$\CC$CDM model (cf. more details in section \ref{sec:num_str}). On
the other hand, neither the value of $\rL$ nor of the perturbation
in this variable are expected to play an important role in the past.
Finally, for reasons that will become apparent later, it is
important to restrict our discussion to models wherein the time
variation of $G$ is very small, as this may also affect the previous
consideration on the transfer function (and on top of this there are
important bounds from primordial nucleosynthesis to be satisfied,
see section \ref{sec:num_nuc}).

After insuring that these conditions are fulfilled by the admitted
class of variable $G$ models, we trust that the $k$-dependence of
the transfer function for these models should be the same as that
for the $\CC$CDM model. Combining this fact with the above proven
scale independence of the late time evolution of the coupled set of
perturbations for $\delta\rL$ and $\delta G$, we reasonably infer
that the matter power spectrum of a model with variable $\rL$ and
slowly variable $G$ -- and with self-conserved matter components --
must generally have the very same spectral shape as that of the
$\CC$CDM model. Fortunately, in spite of their shape invariance,
such models can still be constrained and distinguished from the
standard cosmological model by means of the spectrum amplitude (i.e.
from the normalization of the matter fluctuation power spectrum). We
shall further elaborate on these points in
section~\ref{sec:numerical}, where a concrete model exhibiting these
properties will be analytically and numerically analyzed.

Let us now retake our analysis of cosmic perturbations by showing
that the perturbations in $G$ are actually tightly linked to the
perturbations in $\rL$. This feature is another consequence of the
aforementioned scale invariance of the late time evolution of the
perturbations. Indeed, from (\ref{fin5}) and the neglect of the
velocity perturbations (\ref{theta2}), we obtain:
\begin{equation}
\delta_{\Lambda}\equiv\frac{\delta\rho_{\Lambda}}{\rho_{\Lambda}}=-\frac{\delta
G}{G} \label{prime}\,.
\end{equation}
It follows that the two kind of perturbations are not ultimately
independent. Furthermore, using Eq.\,(\ref{prime}), a
straightforward calculation from (\ref{fin4}) renders the simple
result:
\begin{equation}
\delta_m\equiv \frac{\delta\rho_m}{\rho_m}=-\frac{\dot{\delta
G}}{\dot{G}}=-\frac{{(\delta G)'}}{{G'}}\,. \label{diff2}
\end{equation}
{The last two equations show clearly that the perturbations in $G$
and $\rL$ may not be consistently set to zero, since they will be
generated even if they are assumed to vanish at some initial instant
of time.\footnote{This is analogous to what happens in models with
self-conserved DE, where DE perturbations cannot be consistently
neglected \cite{Kunz:2006wc}}} From (\ref{fin2}), and using the
matter conservation (\ref{standardconserv}), it is easy to see that
\begin{equation}
\hat{h}=2\dot{\delta}_m\,.
\end{equation}
We can now use this last expression and Eq.\,(\ref{fin1}) to produce
a second order differential equation for $\delta_m$. Using again
differentiation with respect to the scale factor, we find
\begin{eqnarray}
\delta_m''+A(a)\delta'_m=B(a)\left(\delta_m+\frac{\delta
G}{G}\right)\,,\label{diff1}
\end{eqnarray}
where we have defined:
\begin{eqnarray}
&&A=\frac{3}{a}+\frac{H'(a)}{H(a)}\\
&&B=\frac{3}{2a^2}\tilde{\Omega}_m(a)\\
&&\tilde{\Omega}_m(a)=\frac{\rho_m(a)}{\rho_c(a)}=\frac{8\pi
G(a)}{3H^2(a)}\rho_m(a)\,. \label{tild}
\end{eqnarray}
Here, $\tilde{\Omega}_m(a)$ is the ``instantaneous'' normalized
matter density at the time where the scale factor is $a=a(t)$. It is
also convenient to define the corresponding instantaneous normalized
CC density $\tilde{\Omega}_\CC(a)={\rL(a)}/{\rho_c(a)}$, and it is
easy to check that the following sum rule
\begin{equation}\label{tildeOmegas}
\tilde{\Om}_m(z)+\tilde{\Om}_\CC(z)=1
\end{equation}
is satisfied for all $z$. Clearly, the sum rule is a simple but
convenient rephrasing of Eq.\,(\ref{Friedmann}).

If we would neglect the perturbations in $G$ ($\delta G\sim0$) and
hence $\delta\rL\sim 0$ too, owing to Eq.\,(\ref{prime}), it is
immediate to see that Eq.\,(\ref{diff1}) would reduce to the
standard differential equation\,\cite{Dodelson} for the growth
factor for DE models with self-conserved matter under the assumption
of negligible DE perturbations, i.e. explicitly Eq.(90) of
Ref.\,\cite{LXCDM08}. In particular, if we take for $H$ the standard
form (\ref{Friedmann}) with $\rL=$const., we arrive to the equation
for the $\CC$CDM model. However, Eq.\,(\ref{diff1}) with $\delta
G\neq0$ tells us precisely how to take into account the effect of
non-vanishing perturbations in the gravitational constant $G$. We
could now solve the coupled system formed by (\ref{diff1}) and
(\ref{diff2}), giving initial conditions for $\delta'_m$, $\delta_m$
and $\delta G$, or use those two equations to get a third order
differential equation for $\delta_m$, in which case we would need to
give the initial condition for $\delta''_m$ instead. This seems to
be the natural thing to do, because it involves boundary conditions
on $\delta_m$ and its derivatives only, and does not mix with the
boundary conditions on other variables.

In order to get the aforesaid third order differential equation for
$\delta_m$, we have to differentiate Eq.\,(\ref{diff1}) and use this
same equation to eliminate $\delta G/G$. Using also (\ref{diff2}),
we obtain
\begin{eqnarray}
\left(\frac{\delta G}{G}\right)'=\frac{\delta
G'}{G}-\frac{G'}{G}\frac{\delta
G}{G}=-\frac{G'}{G}\,\delta_m-\frac{G'}{G}\left(\frac{\delta_m''+A\,\delta_m'}{B}-\delta_m\right)
=-\frac{G'}{G\,B}\left(\delta_m''+A\,\delta_m'\right)\,,
\end{eqnarray}
and hence we finally arrive at the desired third order equation:
\begin{equation}\label{thirdorder}
\delta_m'''+\left(A-\frac{B'}{B}+\frac{G'}{G}\right)\delta_m''+\left[A'-B+A\left(\frac{G'}{G}-\frac{B'}{B}
\right)\right]\delta_m'=0\,.
\end{equation}
Now we have to compute the coefficients of this equation. From the
definition of $\tilde{\Omega}_m$, Eq.~(\ref{tild}), we get the
useful relation:
\begin{equation}
\frac{\tilde{\Omega}_m'}{\tilde{\Omega}_m}=\frac{G'}{G}-2\frac{H'}{H}-\frac{3}{a}\,.\label{primimp}
\end{equation}
On the other hand, differentiating Friedmann's equation
(\ref{Friedmann}),
\begin{equation}
2HH'=\frac{8\pi}{3}[G'(\rho_m+\rho_\CC)+G(\rho_m'+\rho_\CC')=\frac{8\pi
G}{3}\rho_m'=-\frac{8\pi G}{a}\rho_m\,,
\end{equation}
where in the last two steps we have taken into account the
energy-conservation law (\ref{Bianchi}) and
(\ref{solstandardconserv}) for $\alpha_m=3$ (matter epoch).
Combining this last equation with the definition of
$\tilde{\Omega}_m$, (\ref{tild}), we get:
\begin{equation}
\frac{H'}{H}=-\frac{3}{2a}\tilde{\Omega}_m\,.\label{segimp}
\end{equation}
Finally, using (\ref{primimp}) and (\ref{segimp}) {we can rewrite
our third order differential equation (\ref{thirdorder}) in terms of
$\tilde{\Omega}_m$ as}:
\begin{equation}
\delta_m'''+\frac{1}{2}\left(16-9\tilde{\Omega}_m\right)\frac{\delta_m''}{a}+\frac{3}{2}\left(8-11\tilde{
\Omega}_m+3\tilde{\Omega}_m^2-a\,\tilde{\Omega}_m'\right)\frac{\delta'_m}{a^2}=0\,.\label{supeq}
\end{equation}
Let us stress that this equation is valid for \emph{any} model with
variable G and $\rL$ in which matter is covariantly conserved, i.e.
models of the type iv) in section \ref{sec:GenericModels}. However,
a very particular case where it should give the correct results is
for the $\CC$CDM and CDM models because matter is conserved and the
equation (\ref{Bianchi}) is trivially satisfied. For the CDM model,
for instance, $\tilde{\Omega}_m=1$ and the equation reduces to:
\begin{equation}
\delta_m'''+\frac{7}{2}\frac{\delta_m''}{a}=0\,,
\end{equation}
which has the general solution:
\begin{equation}
\delta_m=C_1 a + C_2 + C_3 a^{-3/2}\,.
\end{equation}
Barring the constant and decaying modes, which we can neglect, the
relevant solution is the growing mode $\delta_m\propto a$, which is
linear in the scale factor, as expected. {Note that the setting
$\tilde{\Omega}_m=1$ implies that there is no CC term, and from
Eq.\,(\ref{Bianchi}) we infer that $G$ must be strictly constant in
such case. Finally, Eq.\,(\ref{prime}) entails that $\delta G$ is
then also constant, and this constant can be set to zero through a
redefinition of $G$.} Let us also remark that, in this limit, one
does not expect to recover the perturbative analysis of scenario ii)
in section \ref{sec:GenericModels}, i.e. the results obtained in
\cite{Fabris:2006gt}. The reason is that although in the last
reference there are no perturbations on $G$, matter and vacuum are
exchanging energy. Therefore the underlying physics is completely
different and there is in general no simple connection between these
two kind of models.

We can use the general equation (\ref{supeq}) to study the
perturbations in any model with variable $G$ and $\rL$ in which
matter is conserved. In the next section, we consider a well
motivated non-trivial example.

\section{The class of vacuum power series models in $H$}\label{sec:QFTvac}

As we have mentioned in the introduction, there have been many
attempts to envisage a dynamical cosmological term
$\rL(t)$\,\cite{oldvarCC1}. However, in most cases the approach is
purely phenomenological, with no reference to a potential connection
with fundamental physics, via QFT or string theory. On the other
hand, there are models wherein there is such a possible connection.
Obviously, this represents an additional motivation for their study.
Consider the class of models in which the CC {evolves as a power
series in} the Hubble rate:
\begin{equation}\label{GeneralPS}
\rL(H)=n_{0}+n_1H+n_{2}H^{2}+...\,,
\end{equation}
where $n_{i}\,\,(i=0,1,2,...)$ are dimensional coefficients (except
$n_4$, which is dimensionless). The background solution for this
cosmological model up to $i=2$ can be found in \cite{BPS09}. The
higher order terms in (\ref{GeneralPS}), made out of powers of $H$
larger than $2$, are phenomenologically irrelevant at the present
time and will not be discussed. Besides, let us stress that from the
fundamental point of view of QFT in curved space-time, the general
covariance of the effective action can only admit even powers of the
expansion rate\cite{ShapSol02,CCfit,ShapiroEA}. So the coefficient
$n_1$ of the linear term in $H$ cannot be related to the properties
of the effective action, but to some phenomenological parameter of
the theory (e.g. the viscosity of the DE fluid) which is not part of
the fundamental principles. If we dispense with this kind of
phenomenological coefficients, we obtain the subclass of models of
the form
\begin{equation}\label{RGlaw}
\rL(H)=n_{0}+n_{2}H^{2}\,,
\end{equation}
which have been advocated as scenarios in which the CC evolution can
be linked to the  renormalization group (RG) running in
QFT\,\cite{ShapSol02,ShapSol09a}. Notice that the coefficient $n_2$
has dimension of an effective mass squared, $n_2=M_{\rm eff}^2$. We
shall further comment on it below.

The form (\ref{RGlaw}) is indeed crucially different from just
considering that the vacuum energy is proportional to $H^2$, in the
sense that Eq.\,(\ref{RGlaw}) is an ``affine quadratic law'' (i.e.
$n_0\neq 0$). While the pure $H^2$ law is not favored by the
experimental data\,\cite{BPS09}, the affine version has been
recently tested in a framework where $G$ is constant, specifically
in the framework of the scenario ii) in section
\ref{sec:GenericModels}, and it was found that it can provide a fit
{to the current observational data of similar quality as the
$\CC$CDM one -- see \cite{BPS09} for details.

Consider now the size of the $n_2$ coefficient in
Eq.\,(\ref{RGlaw}). In its absence, the the CC is strictly constant
and $n_0$ just coincides with the current value $\rLo$. However, in
the presence of the $H^2$ correction, the boundary condition at
$H=H_0$ becomes $\rLo=n_0+n_2\,H_0^2$. If the additional term is to
play a significant role it should not be negligible as compared to
$n_0$, and as the latter is the leading term it must still be of
order $\sim\rLo$. It follows that the effective mass $M_{\rm eff}$
associated to the coefficient $n_2$ cannot be small, but actually
very large -- specifically, of order of a Grand Unified Theory scale
(see below). This also explains why no other even power of $H$ can
play any significant role in the series expansion (\ref{GeneralPS})
at any stage of the cosmological history below a typical GUT scale.
The upshot is that, in practice, the evolution law (\ref{RGlaw}) is
the leading form throughout all relevant cosmic epochs (radiation
dominated, matter dominated and late CC dominated epochs).

{Likewise}, the possibility that the vacuum energy could be evolving
linearly with $H$ -- i.e. as if $n_0=n_2=0$ in
Eq.\,(\ref{GeneralPS}) -- has also been addressed in the literature
and can be motivated through a possible connection of cosmology with
the QCD scale of strong
interactions\,\cite{Schu02,Volov09,Urban09b}. However, as we have
said, this option is not what one would expect from the the general
covariance of the effective action. Actually, the confrontation of
the purely linear model $H$ with the data does not seem to support
it\,\cite{BPS09,Borges:2007bh}, and therefore the linear term alone
is unfavored. However, it could perhaps enter as a phenomenological
term in a general power series vacuum model of the form
(\ref{GeneralPS}) -- {a possibility which is currently under
study\,\cite{BGPS}.} For some alternative recent models with
variable cosmological parameters, see e.g. \cite{Alter09}.

In the following, we focus on the quantum field vacuum model of the
form (\ref{RGlaw}). It is particularly convenient to rewrite the
coefficients of this equation as follows:
$n_0=\rLo-3\nu\,M_P^2\,H_0^2/(8\pi)$ and $n_2=3\nu\,M_P^2/(8\pi)$.
Therefore, the CC evolution law reads
\begin{equation}\label{RGlaw2}
 \rL(H)=\rLo+ \frac{3\nu}{8\pi}\,M_P^2\,(H^{2}-H_0^2)\,.
\end{equation}
Here $\nu$ is a small coefficient ($|\nu|\ll 1$) and $M_P$ is the
Planck mass; it defines the current value of Newton's constant:
$G_0=1/M_P^2$. Clearly, the vacuum energy density (\ref{RGlaw2}) is
normalized to the present value, i.e.
\begin{equation}\label{rLo}
\rL(H_0)=\rLo\equiv\frac{3}{8\pi}\,\OLo\,H^{2}_{0}\,M_P^2\,,
\end{equation}
where experimentally $\OLo\simeq 0.7$. The above parametrization
satisfies the aforementioned condition that $n_0$ is the leading
term and is of order $\rLo$, and at the same time the correction
term is of order $M^2_{\rm eff}\,H^2$, with $M_{\rm eff}\sim
\sqrt{\nu}\,M_P$ a large mass, even if $\nu$ is as small as, say,
$|\nu|\sim 10^{-3}$ or less.

It is now convenient to define a new set of cosmological energy
densities normalized with respect to the \emph{current} critical
density $\rco=3H_0^2/(8\pi\,G_0)$:
\begin{eqnarray}
\Omega_i(z)&\equiv&\frac{\rho_i(z)}{\rco}=\frac{E^2(z)}{g(z)}\,\tilde{\Om}_i(z)\
\ \ \ (i=m,\CC)\,,\label{tildom}
\end{eqnarray}
where we have introduced the ratios
\begin{equation}\label{Eg}
E(z)\equiv\frac{H(z)}{H_0}=\sqrt{g(z)}\,\left[\Om_m(z)+\OL(z)\right]^{1/2}\,,
\ \ \ g(z)\equiv\frac{G(z)}{G_0}\,,
\end{equation}
with $G(z)$ Newton's constant at redshift $z$. Notice that, in Eq.\,
(\ref{tildom}), we have explicitly related the new parameters
$\Omega_i(z)$ with the old ones $\tilde{\Om}_i(z)$, the latter being
defined in Eqs.\,(\ref{tild}),(\ref{tildeOmegas}). Obviously, they
all coincide at $z=0$ with the normalized current densities
$\Om_i^0$, i.e. $\Omega_i(0)=\tilde{\Omega}_i(0)=\Omega_i^0$.

Clearly, the coefficient $\nu$ in Eq.\,(\ref{RGlaw2}) measures the
amount of running of the CC. For any given $\nu$, we can compare the
value of the dynamical CC term at a cosmic epoch characterized by
the expansion rate $H$ -- or by the redshift $z$ -- with the current
value (\ref{rLo}). The relative correction can be conveniently
expressed as follows:
\begin{equation}\label{DeltarLo}
\Delta\OL(z)\equiv\frac{\OL(z)-\OLo}{\OLo}=\frac{\nu}{\OLo}\left[E^2(z)-1\right]\,,
\end{equation}
Of course $G(0)=G_0$. Moreover, since $g=1$ for $\nu=0$, it follows
that for small $\nu$, $g(z)$ deviates little from $1$, namely
$g(z)=1+{\cal O}(\nu)$. Thus, expanding to order $\nu$ in the matter
epoch, it is easy to show from the previous equations that
\begin{equation}\label{deltarhoCC}
\Delta\OL(z)\simeq
\nu\,\frac{\Om_m^0}{\OLo}\,\left[(1+z)^3-1\right]\,,
\end{equation}
where $g(z)\sim 1$ to this order. If we look back to relatively
recent past epochs, e.g. exploring redshifts $z={\cal O}(1)$
relevant for Type Ia supernovae measurements, we see that
$\Delta\OL(z)$ can be of order of a few times $\,\nu$. For example,
for $z=1.5$ and $z=2$,  we have $\Delta\OL(1.5)\simeq 6\,\nu$ and
$\Delta\OL(2)\simeq 11\,\nu$ respectively, assuming $\OM^0=0.3$. The
correction is thus guaranteed to be small, as desired, but is not
necessarily negligible. We shall see, in the next sections, the
potential implications on important observables, which will actually
put tight bounds on the value of $\nu$.

Although the above parametrization of the CC running can be purely
phenomenological, let us recall that the dimensionless coefficient
$\nu$ can be interpreted, in more fundamental terms, within the
context of QFT in curved-space time; specifically it is proportional
to the ``$\beta$-function'' for the RG running of the CC term. The
predicted value in this QFT framework is
\cite{ShapSol02,CCfit,SSS04,Fossil07}
\begin{equation}\label{nu}
\nu=\frac{\sigma}{12\pi}\,\frac{M^2}{M_P^2}\,,
\end{equation}
where $M$ is an effective mass parameter, representing the average
mass of the heavy particles of the Grand Unified Theory responsible
for the CC running through quantum effects (after taking into
account the multiplicities of the various species of particles).
Obviously, $M\sim M_{\rm eff}$. Since $\sigma=\pm 1$ (depending on
whether bosons or fermions dominate in the loop contributions), the
coefficient $\nu$ can be positive or negative, but $|\nu|$ is
naturally predicted to be smaller than one. For instance, if GUT
fields with masses $M_i$ near $M_P$ do contribute, then
$|\nu|\lesssim 1/(12\pi)\simeq 2.6\times 10^{-2}$, but we expect it
to be even smaller in practice because the usual GUT scales are not
that close to $M_P$. By counting particle multiplicities in a
typical GUT, a natural estimate lies in the range
$\nu=10^{-5}-10^{-3}$ (see \,\cite{SSS04,Fossil07} for details).

{In the next section, we shall concentrate on a specific running QFT
vacuum model of the type (\ref{RGlaw2}) where the vacuum energy and
the gravitational constant vary simultaneously in accordance to the
Bianchi identity (\ref{Bianchi}). We shall also perform a detailed
numerical analysis of our results.}

\section{Application: running QFT vacuum
model with variable G}\label{sec:numerical}

In this section, we consider a CC running vacuum model in which
$\rL=\rL(H)$ depends on the Hubble rate through the affine quadratic
law (\ref{RGlaw2}) and in which matter is separately conserved --
the time variation of the CC being compensated by that of the
Newton's coupling, $G=G(H)$. This setup corresponds to the scenario
iv), as defined in section~\ref{sec:GenericModels} and was first
considered at the background level with alternative motivations in
Refs.\,\cite{SSS04,Fossil07}. The novelty here is to consider the
detailed treatment of the cosmic perturbations in that scenario.
Namely, we apply to it the general treatment of cosmic perturbations
with variable $\rL$ and $G$ developed in section
\ref{sec:general_perturb}. In this way, we can study the growth of
matter density perturbations and obtain a LSS bound on the basic
parameter $\nu$, the one that controls the variation of $G$ and
$\rL$. Actually, the tightly bounded region will appear in the
$(\nu,\OM^0)$ plane.

Specifically, we will constrain the parameter $\nu$ (\ref{nu}) by
requiring that the amount of growth of matter perturbations in our
model does not deviate too much from the value deduced from the
observations of the number density of local clusters. In fact, let
us recall that these observations allow to set the normalization of
the matter power spectrum\,\cite{Pierpaoli:2000ip,Henry:2008cg}, and
hence fix its amplitude. In view of the shape independence of the
power spectrum for the models under study, which we have discussed
in section \ref{sec:general_perturb}, the truly relevant parameter
to constraint our model is the spectral amplitude\,\cite{Dodelson}.
As we shall see, the constraints obtained in this way are in very
good agreement with those arising from primordial nucleosynthesis.
Throughout this section, we will be using the scale factor and the
cosmological redshift $z=(1-a)/a$ interchangeably.

The cosmological evolution of our model is determined by the
Friedmann equation (\ref{Friedmann}), the Bianchi identity
(\ref{Bianchi}) and the RG law for the CC (\ref{RGlaw2}). Using the
normalized density parameters defined in Eq.\,(\ref{tildom}), the
basic cosmological equations can be formulated as
\begin{eqnarray}\label{System1}
&&E^2(z)=g(z)\left[\OM(z)+\OL(z)\right]\label{Friedmann2}\,,\\
&&(\OM+\OL)dg+g\,d\Om_\CC=0\,,\label{bianchi2}\\
&& \OL(z)=\OLo+\nu\left[E^2(z)-1\right]\,, \label{RGlaw3}\\
&&\OM(z)=\Om_m^0\,(1+z)^{3(1+\omega_m)}\,,\label{conservOm}
\end{eqnarray}
where the last equation just reflects the covariant conservation of
matter, i.e. it is a rephrasing of Eq.(\ref{solstandardconserv}).
Solving the remaining system for the function $g=g(H)$, it is easy
to arrive at
\begin{equation}\label{GH}
g(H)\equiv
\frac{G(H)}{G_0}=\frac{1}{1+\nu\,\ln\left(H^2/H_0^2\right)}\,.
\end{equation}
We confirm that $g(H)$ depends also on the parameter $\nu$ and that,
to first order, we have $g=1+{\cal O}(\nu)$. Thus, in this model,
$\nu$ plays also the role of the $\beta$-function for the RG running
of $G$. Compared to the quadratic running of the CC with the Hubble
rate, indicated in Eq.\,(\ref{RGlaw3}), the running of $G$ with $H$
is indeed very slow, it is just logarithmic. The functions $g(z)$
and $\OL(z)$ cannot be determined explicitly in an analytic form.
Nevertheless, it is possible to derive an implicit equation for
$g(z)$ by combining (\ref{GH}) with (\ref{System1}) and
(\ref{RGlaw3}). For the matter-dominated epoch, the final result
reads:
\begin{equation}\label{implicit}
\frac{1}{g(z)}-1+\nu\,\ln\left[\frac{1}{g(z)}-\nu\right]=
\nu\,\ln\left[\OM(z)+\OLo-\nu\right]\,,
\end{equation}
with $\OM(z)=\Om_m^0\,(1+z)^{3}$. The CC density follows from
(\ref{RGlaw3}), and is given as a function of $g(z)$:
\begin{equation}\label{CCz}
\OL(z)=\frac{\OLo+\nu\left[\OM(z)\,g(z)-1\right]}{1-\nu\,\,g(z)}\,.
\end{equation}
\begin{figure}
\includegraphics[width=0.49\textwidth]{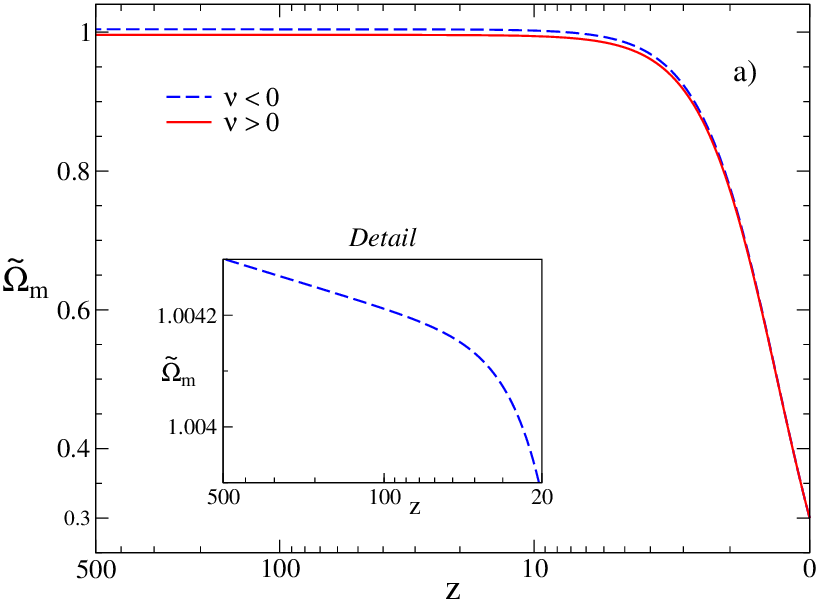}
\includegraphics[width=0.49\textwidth]{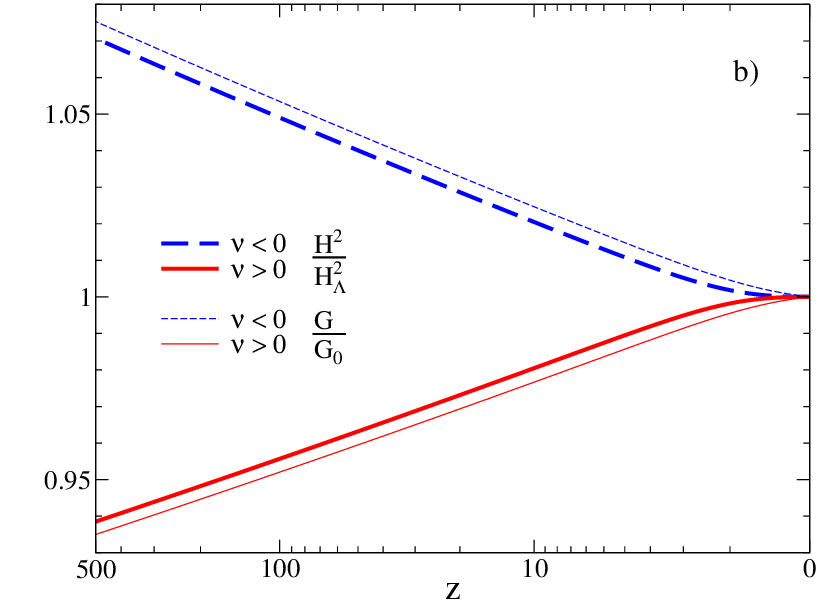}\vspace{0.3cm}
\caption{Evolution of different background quantities for the QFT
cosmological model with running $\CC$ and $G$, in terms of the
redshift $z$. We take $\Omo=0.3$ and consider both a positive
($\nu=+4\times 10^{-3}$, red solid lines) and a negative value
($\nu=-4\times 10^{-3}$, blue dashed lines) for the parameter $\nu$,
which controls the variation of $G$ and $\CC$; a) the matter density
fraction, $\tilde{\Om}_m(z)$ (\ref{tild}), rapidly approaches unity
in the past, meaning that for high redshifts
$\tilde{\Omega}_{\CC}(z)$ is negligible and our model closely
resembles a CDM (for which $\tilde{\Om}_m(z)$ is \emph{exactly} 1);
b) evolution of $G(z)$ (thin lines) and $H^2(z)$ (thick lines),
normalized to the $\CC$CDM values, showing that the departure from
the standard cosmological model is small. The evolution of both
quantities is related by Eq.~(\ref{relHg}).}\label{fig1}
\end{figure}
As a simple check, we can see that Eq.\, (\ref{implicit}) is
satisfied at $z=0$ for any value of $\nu$, using $g(0)=1$ and the
sum rule $\Omega_m^0+\Omega_{\CC}^0=1$ for flat space. Then from
(\ref{CCz}) we immediately obtain $\OL(0)=\OLo$, as expected.

In Fig.~\ref{fig1}, we show the evolution of different background
quantities in this model in terms of the redshift $z$. For the
plots, we have used $\Omo=0.3$ and both a positive
($\nu=+4\times10^{-3}$, red solid line) and a negative value
($\nu=-4\times10^{-3}$, blue dashed line) for the parameter $\nu$.
As we will see later in this section, according to LSS and
nucleosynthesis considerations, we expect $\nu$ to be (at most) of
order $10^{-3}$, so these are reasonable values. In
Fig.~\ref{fig1}a, we plot the evolution of the matter density
fraction, $\tilde{\Om}_m(z)$ (\ref{tild}), which rapidly approaches
unity in the past. This means that $\tilde{\Omega}_{\CC}(z)$ is
negligible for high redshifts -- recall the sum rule
(\ref{tildeOmegas}) -- so that in this regime our model closely
resembles a CDM model (for which we would have $\tilde{\Om}_m(z)$
\emph{exactly} 1.) This could have been anticipated from
(\ref{RGlaw3}), which far in the past reads:
\begin{equation}
\OL(z)\simeq\nu \,E^2(z)\,.
\end{equation}
Using this expression in (\ref{tildom}), we obtain the corresponding
asymptotic value of $\tilde{\Om}_\CC$ in the past:
\begin{equation}
\tilde{\Om}_\CC(z)= \frac{g(z)}{E^2(z)}\,\Om_\CC(z)\simeq \nu
g(z)\sim\mathcal{O}(\nu)\ll 1 \ \ \ \ \ \ \ (z\gg 1)\,,
\label{OLnegli}
\end{equation}
since indeed $g(z)\sim\mathcal{O}(1)$, as can be confirmed from
Fig.~\ref{fig1}b. Equation (\ref{OLnegli}) tells us that, for any
reasonable value of $\nu$, the contribution of the running CC in the
past will be unimportant. Nevertheless, as seen in the detail frame
of Fig.~\ref{fig1}a, $\tilde{\Om}_m(z)$ in our model is indeed not
exactly one, nor really constant. Note that for $\nu<0$ the CC
density decreases as we go into the past (the opposite is true for
$\nu>0$) until it eventually gets negative. Therefore, our model can
accommodate either originally positive or negative values for $\CC$,
thanks to the running nature of this quantity. Notice that the
previous results can be viewed as being a consequence of the fact
that, for flat space, the tilded normalized densities satisfy the
sum rule (\ref{tildeOmegas}).

In Fig.~\ref{fig1}b we show the evolution of $g(z)$ and
$H^2(z)/H^2_\CC(z)$ where $H_\CC$ is the (flat) $\CC$CDM Hubble
function, simply given by:
\begin{equation}
H_\CC^2=H_0^2\left[\OM(z)+\OLo\right]=H_0^2\left[\OM(1+z)^3+\OLo\right]\,.
\end{equation}
Taking into account (\ref{Friedmann2}) and (\ref{CCz}) we obtain, to
order $\nu$,
\begin{equation}
\frac{H^2(z)}{H_\CC^2}=
g(z)\,\frac{\OM(z)+\OL(z)}{\OM(z)+\OLo}\simeq
g(z)\,\left(1+\nu\frac{\OM(z)-\OM^0}{\OM(z)+\OLo}\right)\,. \ \ \ \ \
\label{relHg}
\end{equation}
Therefore the evolutions of $H^2(z)/H_\CC^2$ and $g(z)$ are expected
to be very similar, as indeed  shown in Fig.~\ref{fig1}b.
Furthermore, both quantities stay close to 1, so the deviation from
the standard $\CC$CDM evolution is reasonably small, although it
maybe large enough so as to be detected in a future generation of
precision cosmology experiments. For instance, for the values of
$\nu$ and $\Omo$ that we are considering, the relative deviations of
$\OL(z)$, $g(z)$ and $H^2(z)$ at $z=2$ with respect to the standard
model values are (approximately) $4\%$,  $1\%$ and  $0.5\%$,
respectively.

\subsection{Constraints from large-scale structure}\label{sec:num_str}

Let us now move to the detailed study of the growth of matter
density perturbations in our model, and further elaborate on some
important issues raised in section \ref{sec:general_perturb}. The
matter power spectrum for a sufficiently recent value of the scale
factor $(a\gg a_{\rm eq})$ can be written as:
\begin{equation}
P(k,a)\equiv\left|\delta_m(k,a)\right|^2=\mathcal{A}k^nT^2(k)D^2(k,a)\,.\label{spec}
\end{equation}
Here $\mathcal{A}$ is a normalization coefficient; $n$ is the scalar
spectral index (which gives us the shape of the primordial spectrum;
e.g. $n=1$ if we assume a Harrison-Zel'dovich spectrum); $T(k)$ is
the transfer function, which does not depend on the initial
conditions but only on the physics of the microscopic constituents;
i.e. it encodes the modifications of the primordial spectrum that
arise when taking into account the dynamical properties of the
particles and their interactions. The $T(k)$ function receives
non-trivial contributions only from the evolution of the different
(comoving) wavelenghts $\lambda\sim 1/k$ in the epochs of radiation,
horizon crossing and radiation/matter equality, i.e. it reflects the
$k$-dependent features that occur at early times when $a$ moves from
$a\ll a_{\rm eq}$ to $a\gg a_{\rm eq}$. In general, the form of
$T(k)$ depends on the cosmological model under consideration, and
determines to a large extent the final shape of the spectrum.
However, for the late time evolution ($a\gg a_{\rm eq}$), there
might be more distortions of the spectrum in models that depart
significantly from the standard $\CC$CDM scenario (for which there
are no further $k$-dependence beyond that already encoded in the
transfer function). These additional, model-dependent, effects are
reflected in the $k$-dependence of the growth factor $D(k,a)$ in
Eq.\,(\ref{spec}). A dependence of this sort would appear e.g. in
models of type ii) in section \ref{sec:GenericModels} where the CC
decays into matter, as it was previously studied
in\,\cite{Fabris:2006gt}. However, our claim (formulated in section
\ref{sec:general_perturb}) is that this is not the case for the
variable $CC$ and $G$ models under consideration, i.e. models of
type iv) in section \ref{sec:GenericModels} with self-conservation
of matter, provided that $G$ is a slowly varying function of time.
Obviously this is so for the model studied in the previous section,
where $G$ varies logarithmically with the expansion of the universe,
see Eq.\,(\ref{GH}). For these models, and of course also for the
standard $\CC$CDM model, the growth factor is independent of $k$ and
can be written as
\begin{equation}\label{growthfactor}
D(a)=\frac{\delta_m(a)}{\delta_{\rm CDM}(a_0)}\,,
\end{equation}
where $\delta_{\rm CDM}(a_0)$ is the present matter density contrast
in a pure cold dark matter (CDM) scenario, taken as a fiducial model.

\subsubsection{More on the transfer function for variable $\rL$ and $G$
models.}\label{sec:moreTransfer}

A few additional comments on the transfer function are now in order.
As previously commented, we expect the transfer function in our
model to be the same as for the $\CC$CDM. First of all, let us try
to better justify this expectation, which implies that the shape of
the matter power spectrum in our model coincides with that of a
$\CC$CDM with the same value of $\Omo$. Then we show that we can
still constrain our model by means of the growth factor, obtaining
bounds in very good agreement with those arising from primordial
nucleosynthesis.

The scale dependence of the transfer function for CDM models is
different for large scales (small $k$'s), where $T(k)=1$, and small
scales, where it asymptotes to $\ln (k)/k^2$\,\cite{Dodelson}. The
turnaround occurs at the value $k=k_{\rm eq}$, corresponding exactly
to the scale that enters the Hubble horizon ($H^{-1}$) at the moment
of matter-radiation equality, i.e. at the point $a=a_{\rm eq}$ which
satisfies $\OM(a_{eq})=\OR(a_{eq})$, hence
\begin{equation}\label{aeq}
a_{\rm eq}=\frac{\ORo}{\Omo}\,,
\end{equation}
where $\,\ORo\sim10^{-4}$ is the present density of radiation. The
modes that enter the horizon at the equality time, with comoving
wavelength $\lambda_{\rm eq}$, have a physical wavelength that
follows upon multiplication with the scale factor (\ref{aeq}), i.e.
$\lambda_{\rm eq}\,a_{\rm eq}=1/H(a_{\rm eq})$, or, equivalently,
\begin{equation}
k_{\rm eq}=a_{\rm eq}H(a_{\rm eq})\,.
\end{equation}
Obviously, since $H$ decreases with time the perturbations with
wavelenght shorter than $\lambda_{\rm eq}$ (i.e. those with with
$k>k_{\rm eq}$) will enter the Hubble horizon before the
matter-radiation equality, i.e. in the radiation era ($t<t_{\rm
eq}$). From this moment until $t=t_{\rm eq}$, the growth of
inhomogeneities in the cold DM component becomes suppressed because
the expansion rate during the radiation epoch is faster than the
characteristic collapsing rate of the CDM. As a result, the modes in
the radiation epoch can grow at most logarithmically with the scale
factor. Only after the cold component begins to dominate ($t>t_{\rm
eq}$) the amplitude of the formerly inhibited modes starts growing
linearly with the scale factor. Finally, as light can only cross
regions smaller than the horizon, the suppression in the radiation
epoch does not affect the large-scale perturbations ($k\ll k_{\rm
eq}$), which enter the horizon in the matter epoch. Such different
behavior of the perturbations according to their entrance in the
horizon before or after the time of equality is the origin of the
characteristic shape of the transfer function for CDM-like models in
the various available parameterizations\,\cite{Dodelson}.

From the previous standard discussion, it is apparent that the shape
of the transfer function depends critically on the value of $k_{\rm
eq}$, which in turn depends on $a_{\rm eq}$ and $H(a_{\rm eq})$. In
the models we are studying, dark matter and radiation are separately
conserved, and therefore the value of $a_{\rm eq}$ will not change
with respect to the standard one, Eq.\, (\ref{aeq}). So the
remaining issue is to clarify if the change of $H(a_{\rm eq})$ in
our model as compared to the corresponding $\CC$CDM value is
significant or not. The answer follows easily from
Eq.\,(\ref{relHg}). At the high redshift where equality of matter
and radiation occurs, $z={\cal O}(10^3)\gg 1$, the function that
accompanies $\nu$ on the \textit{r.h.s.} of that equation is
virtually equal to one, and we are left with
\begin{equation}\label{eq:Heq}
H(a_{\rm eq})\simeq \sqrt{g(a_{\rm eq})\,(1+\nu)}\,H_\CC(a_{\rm eq})\,.
\end{equation}
Here, as before, $H_\CC$ is just the standard $\CC$CDM Hubble rate.
For the maximal values of $\nu$ that we will be considering
($|\nu|\sim {\cal O}(10^{-3})$, see the subsequent sections), it is
easy to see from (\ref{GH}) that
\begin{equation}
\left|\sqrt{g(a_{\rm eq})}-1\right|\simeq |\nu|\,\ln\frac{H(a_{\rm
eq})}{H_0}\sim1\%\,,
\end{equation}
where the expansion rate at the time of equality is $H(a_{\rm
eq})\sim10^{5}\, H_{0}$. We see that the change of $H(a_{\rm eq})$
with respect to $H_\CC(a_{\rm eq})$ is mainly caused by the
variation of $g$ from $1$ at $t=t_{\rm eq}$. However, numerically,
the effect is very small. These results allow us to safely conclude
that the value of $k_{\rm eq}$ expected in our model is essentially
the same as that of the $\CC$CDM model, up to differences of $1\%$
at most. Given that in the $\CC$CDM model the (comoving) wavenumber
at equality is
\begin{equation}\label{eq:keq}
k_{\rm eq}=a_{\rm eq}H_{\CC}(a_{\rm eq})=\sqrt{\frac{2}{\Om_R^0}}\
\Omo H_0\sim 10^{-2}\,h\,\text{Mpc}^{-1}\,,
\end{equation}
we see from this expression that a $1\%$ change in $\sqrt{g(a_{\rm
eq})}$ -- hence in $H(a_{\rm eq})$ -- is equivalent to a $1\%$
change in $\Omo$ in the standard scenario (i.e. with $g=1$). This
variation is too small to be within reach of the present
observations (see the next section), and can be safely neglected.

The final point is that, as argued throughout this section, neither
$\rho_\CC$ nor its perturbations or those of $G$ are important in
the past. Therefore, the evolution of the perturbations both in the
radiation and in the matter-dominated eras (and both in the case of
sub-Hubble and super-Hubble perturbations) remains essentially
unchanged.

All in all, we expect that the transfer function in the model under
consideration is very close to that of a $\CC$CDM with the same
value of $\Omo$. From the line of our argumentation that we have
used, it is not difficulty to convince oneself that this result can
be extended to any model with variable $\CC$ and $G$, and
self-conserved matter, as long as $G(a)$ is not changing too fast.
Adding this property to the fact -- cf.
section~\ref{sec:general_perturb} -- that the late growth of matter
density perturbations in these models does not depend on the
wavenumber, the final robust conclusion is that the matter power
spectrum for models within the scenario iv) of section
~\ref{sec:GenericModels} will present the same shape as in the
standard $\CC$CDM case. Therefore, the spectrum shape will not be
useful to constrain the additional free parameters of the model.
This is in sharp contrast to models within scenario ii), in which
there is an exchange between dark matter and vacuum energy. For
these models there is an explicit dependence on $k$ beyond that of
the transfer function and this produces a late time distortion of
the power spectrum with respect to the $\CC$CDM. As indicated, this
was exemplified in Ref.\,\cite{Fabris:2006gt,Grande:2007wj} for the
case of an evolution law of the type (\ref{RGlaw}).

To summarize, there is a significant difference between the running
QFT vacuum model (\ref{RGlaw}) when studied either in scenario ii)
or when considered in scenario iv). While in Ref.
\cite{Fabris:2006gt,Grande:2007wj} the shape of the spectrum was
used to restrict the parameter $\nu$ for type ii) models, in the
next section we show that for the alternative type iv) models, being
them shape-invariant with respect to the $\CC$CDM model, one can
make use of the amplitude of the power spectrum in order to
constrain the free parameters.

\subsubsection{The spectrum amplitude as a way to constrain running G and $\CC$ models.}
The normalization of the matter power spectrum on scales relevant to
large-scale structure can be performed through different methods
\cite{Bartelmann:1999yn}, e.g. from the microwave-background
anisotropies or by measuring the local variance of galaxy counts
within certain volumes. One of the most robust ways to do it is
through the number density of rich galaxy clusters
\cite{Pierpaoli:2000ip,Henry:2008cg}, which is very sensitive to the
amplitude of the dark matter fluctuations that collapsed to form
them. The typical scales for these fluctuations are of order $10$
Mpc, which are the smallest ones still in the linear part of the
spectrum. The cluster method determines the amplitude of the power
spectrum on just that length scale (corresponding to wavenumbers in
the upper end of those explored with this method). The normalization
is usually phrased in terms of $\sigma_8$, the root mean square mass
fluctuation in spheres with radius $8h^{-1}$ Mpc (i.e. $\sim 10$
Mpc, for $h\simeq 0.7$). By assuming a $\CC$CDM model, these studies
are able to provide constraints in the $\sigma_8$ - $\Omo$ plane. An
important feature is that the results are approximately independent
of the spectrum shape and galaxy bias \cite{Pierpaoli:2000ip}. For
instance, in this last reference, it is found the relation:
\begin{equation}
\sigma_8=\left(0.495^{+0.034}_{-0.037}\right)(\Omo)^{-0.60}\,,
\end{equation}
valid for spatially flat models with a wide range of shapes (in
particular, valid for $0.2\leq\Omo\leq0.8$). When combined with CMB
analyses, cluster studies can determine both $\sigma_8$ and $\Omo$. In a
recent work \cite{Henry:2008cg}, local cluster counts are used in
conjunction with WMAP5 data to find:
\begin{equation}
\Omo=0.30^{+0.03}_{-0.02}\quad (68\%)\,. \label{omcluster}
\end{equation}
We will use this result to constrain our $G$-variable model. In
order to do this, we first define the ``amount of growth'', namely
the square of the growth factor (\ref{growthfactor}) at present,
$D^2(a_0)$, which appears directly in the formula of the power
spectrum, Eq.\,(\ref{spec}). The idea is to compare the amount of
growth in our model model with the amount of growth in the $\CC$CDM
model with $\Omo=0.30$. From the standard expression for the growth
factor in the  $\CC$CDM model\,\cite{Dodelson} one finds that a
$\sim10\%$ variation in $\Omo$ given by (\ref{omcluster}) represents
a $\sim5\%$ change in the amount of growth. As the determination
(\ref{omcluster}) entails only a 1$\si$ constraint, we will be
conservative and allow up to a $10\%$ deviation in the amount of
growth of our model with respect to the $\CC$CDM model. We are thus
asking our model to pass the following ``F-test''
\cite{Grande:2007wj, Ftest}:
\begin{equation}
F=\left|1-\frac{D^2(a_0,\Omo,\nu)}{D^2(a_0,0.3,0)}\right|\leq0.1\,.\label{ftest}
\end{equation}
Notice that $D^2(a_0,0.3,0)$ in the denominator is just the amount
of growth in the  $\CC$CDM for the central value of
(\ref{omcluster}), whereas in the numerator we have the amount of
growth for the model under consideration at a given non-vanishing
value of the relevant parameter $\nu$, with $\,\Omo$ left as a free
parameter. As a result, the constraint (\ref{ftest}) will generate
contours in the $\nu$ - $\Omo$ plane which will define the allowed
region in parameter space.

The admissible values for $\Omo$ should, in principle, be those
compatible with the shape of the spectrum. However, the shape has
been measured by the 2dFGRS and SDSS surveys, existing a significant
difference between their results. On the one hand (assuming a Hubble
parameter $h = 0.72$), the SDSS main galaxy analysis \cite{Teg04}
favored the result
\begin{equation}
\Omo = 0.296 \pm 0.032\,. \label{SDSS}
\end{equation}
Similar values around $\Omo \simeq 0.3$ were found by alternative
analyses of the SDSS catalogue \cite{Pope:2004cc}. On the other
hand, the 2dFGRS collaboration \cite{Cole:2005sx} found a much lower
matter density
\begin{equation}
\Omo = 0.231 \pm 0.021\,, \label{2df}
\end{equation}
obtained from measurements of clustering of blue-selected galaxies.
The inclusion of Luminous Red Galaxies (LRGs) in the SDSS analysis
seems to even increase the discrepancy. For instance, the authors of
\cite{Percival:2006gt} find $\Omo= 0.32 \pm 0.01$, although a lower
density is recovered when restricting the analysis to large scales
($\Omo = 0.22 \pm 0.04\,$ for $\,0.01 < k < 0.06\,h\,{\rm
Mpc}^{-1}$.) It is widely believed that the differences are due to
the scale-dependence of the galaxy bias (which is apparently
stronger for the red galaxies that dominate the SDSS catalogue) or
even to some kind of systematic effect, but the problem has not been
fully settled so far \cite{Sanchez:2007rc}. As an example, the last
result by the SDSS team \cite{Reid:2009xm}, obtained from the
analysis of a LRGs sample (in combination with WMAP5 data), yields
$\Omo = 0.289\pm0.019$, still much larger than (and incompatible
with) the 2dfGRS result, (\ref{2df}), although if we make allowance
for a 10\% gaussian uncertainty in $h$ in the 2dfGRS analysis can
bring both results within $1\si$.

At the end of the day, and taking into account the results from
2dFGRs and SDSS, we think that it would be premature to discard any
value for $\Omo$ in the range $(0.21,\,0.33)$ on the grounds of
structure formation data -- as long as it predicts the right amount
of growth, e.g. by satisfying the F-test (\ref{ftest}). Therefore,
in our analysis, for illustrative purposes, we will consider values
of $\Omo$ between 0.2 and 0.4.

\subsubsection {Numerical results.}

In this numerical section, and in view of the previous
considerations, we wish to determine the set of points in the $\nu$
- $\Omo$ plane for which the amount of growth, determined by the
value of $D^2(a_0)$, in our running $\rL$ and $G$ model deviates
less than $10\%$ from the central ($\Omo=0.3$) $\CC$CDM value in
Eq.\,(\ref{omcluster}). To compute the growth factor in our model,
we evolve the solution $\delta_m(a)$ of the differential
equation~(\ref{supeq}) from an initial value  $a=a_i$ up to the
present moment ($a_0=1$), where $a_i\ll1$ is the scale factor at
some early time, deep into the matter-dominated era but well after
recombination (so that the transfer function regime has already
ended). We will take $a_i=1/501$ (i.e. $z_i=500$), although we have
checked that the results do not the depend significantly on the
specific value. As we have seen in Fig.~\ref{fig1}a, early on at
$a=a_i$ our model is very similar to the CDM model
($\tilde{\Om}_m\simeq1$), for which the matter density perturbations
grow linearly with the scale factor, i.e. $D(a)=a$. Therefore, we
will assume that this is also the case for our running $G$ and $\rL$
model, and take $D(a_i)=a_i$, $D'(a_i)=1, D''(a_i)=0$ as the initial
conditions for the third-order differential equation (\ref{supeq}).

\begin{figure}[t]
\includegraphics[width=0.49\textwidth]{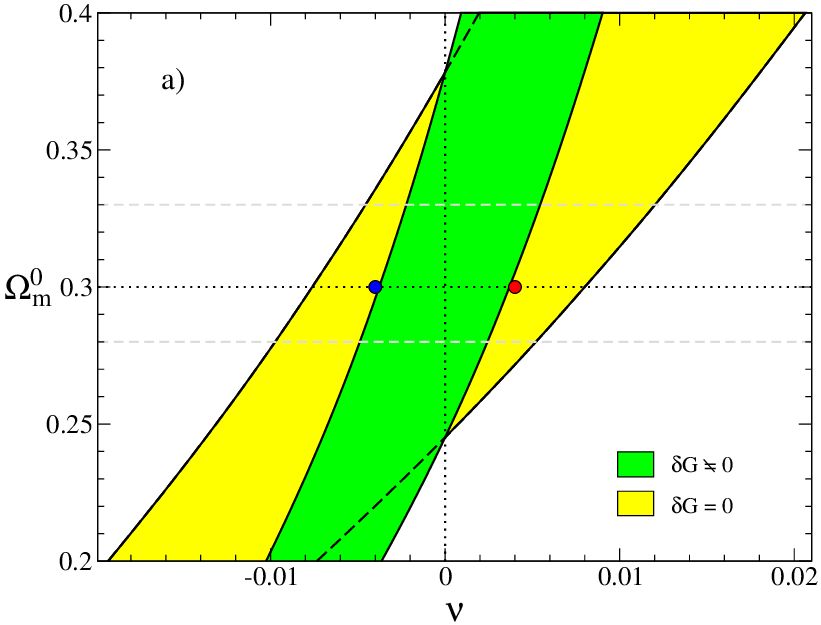}
\includegraphics[width=0.49\textwidth]{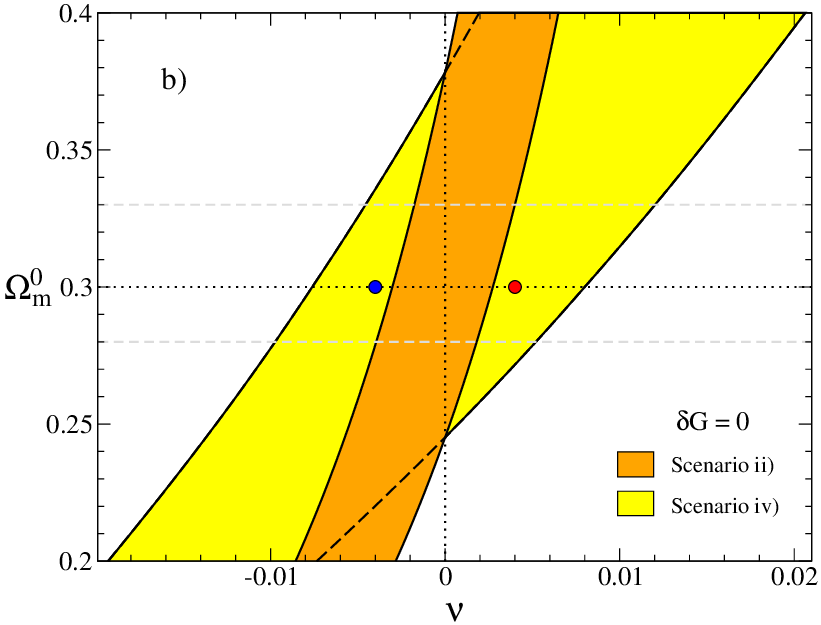}
\caption{Analysis of the parameter space of the running QFT models.
The shaded areas represent the points for which the amount of growth
($D^2(a_0)$) of the model deviates less than 10\% from the $\CC$CDM
value. For the latter, we take $\Omo=0.3$, Eq.\,(\ref{omcluster}),
on the basis of data on the matter power spectrum amplitude; a)
Scenario iv) section \ref{sec:GenericModels} (running $\CC$ and $G$,
self-conserved matter). The green (resp. yellow) band include
(exclude) perturbations in $\rL$ and $G$. With perturbations, the
constraint on the parameter space becomes tighter; b) Comparison
between two scenarios of section \ref{sec:GenericModels} when
perturbations in $\rL$ and $G$ are neglected: scenario iv) (yellow
band) and scenario ii) (orange band). The deviations from the
$\CC$CDM case are larger in scenario ii) owing to the
production/decay of matter from the running $\rL$ at fixed $G$, and
the constraint is correspondingly tighter. The blue and red points
in the plane have coordinates $(\nu=\pm 4\times 10^{-3}, \Omo=0.3)$
-- used in Figs.~\ref{fig1} and \ref{fig3}. The dashed horizontal
lines signal the $1\sigma$ limits on $\OM^0$ from
Eq.\,(\ref{omcluster}).}\label{fig2}.
\end{figure}

The results are shown in Fig.~\ref{fig2}a. The shaded areas
represent the points allowed by our analysis. Specifically, we
compare the case with perturbations in $\rL$ and $G$ (green band)
with the case in which these perturbations are neglected (yellow
band). In the last situation, the growth factor is obtained by
solving Eq.~(\ref{diff1}), although, as discussed in
section~\ref{sec:general_perturb}, it is not possible to neglect
them consistently. The most remarkable conclusion that emerges from
this numerical analysis is that by considering the effect of the
perturbations results in narrower domains, meaning that the
deviations with respect to the standard $\CC$CDM amount of growth
tend to be larger -- which is why the restriction is accordingly
tighter. Therefore, the outcome of the analysis with $\delta G\neq0$
is that, for any $\Omo$ in the range $0.2<\Omo< 0.4$, values of
$|\nu|$ larger than $10^{-2}$ are ruled out by the data on the
number density of local clusters. In particular, this excludes the
\emph{canonical} value $|\nu|=1/12\pi\simeq 0.026$ obtained from the
simplest choice $M=M_P$ in Eq.\,(\ref{nu}). Notice that for $\Omo$
in the narrower ($1\sigma$) interval (\ref{omcluster}), the allowed
values for $|\nu|$ are of order $10^{-3}$ at most. For these values
of $\nu$, our model is on equal footing with the $\CC$CDM as far as
structure formation is concerned.

In Fig.~\ref{fig2}b we compare the results for two of the scenarios
of section \ref{sec:GenericModels}: scenario iv), represented by the
yellow band (this is the scenario we have been analyzing so far,
with variable $\rL$ and $G$ and self-conserved matter; and scenario
ii), indicated by the orange band; for the latter, $G$ is constant
and the CC exchanges energy with matter. In both cases, we are
neglecting the perturbations\footnote{Remember that within scenario
ii), including the perturbations in $\CC$ causes the growth factor
to become scale-dependent \cite{Fabris:2006gt}, $D=D(k,a)$, so the
kind of analysis we are performing here would be no longer
possible.} in $\rL$, which in scenario iv) implies that $\delta G=0$
as well, cf. Eq.\,(\ref{prime}). The effective equation for the
matter density contrast in scenario ii) is the following:
\begin{eqnarray}
&&\ddot{\delta}_m+\left(2H+\frac{\Psi}{\rho_m}\right)\dot{\delta}_m-\left[4\pi G\rho_m-2H\frac{\Psi}{\rho_m}-\frac{d}{dt}\left(\frac{\Psi}{\rho_m}\right)\right]\delta_m=0\,,\\[1ex]
&&\Psi\equiv\dot{\rho}_m+3H\rho_m=-\dot{\rho}_\CC\,.
\end{eqnarray}
This equation, whose primary derivation was performed within the
Newtonian formalism\,\cite{Arcuri:1993pb}, can also can be derived
from the general relativistic treatment of
perturbations\,\cite{LXCDM08}, as explained in \cite{BPS09}. It is
convenient to express it in terms of the scale factor $a$:
\begin{equation}
\delta_m''+\left(\frac{3}{a}+\frac{H'}{H}+\frac{\Psi}{\rho_m H
a}\right)\delta_m'-\left[\frac{3}{2}\tilde{\Om}_m-\frac{2}{H}\frac{\Psi}{\rho_m}-\frac{a}{H}\left(\frac{\Psi}{\rho_m}\right)'\,\right]\frac{\delta_m}{a^2}=0\,.
\end{equation}
For $\Psi=0$, we recover equation (\ref{diff1}) (with $\delta G=0$),
as expected. In Fig.~\ref{fig2}b, we see that the differences in the
amount of growth with respect to the $\CC$CDM case are larger for
scenario ii); the natural interpretation is that this is caused by
the production/decay of matter associated to the time evolution of
$\rL$.

The blue and red points in Figs.~\ref{fig2}a,b correspond to the
values analyzed in Fig.~\ref{fig1}, i.e. $\Omo=0.3$ and $\nu=\pm
4\times 10^{-3}$. We will use again these values to exemplify the
evolution of the matter and Newton's coupling perturbations.
Fig.~\ref{fig3}a shows the growth factor as a function of the
redshift, both when allowing for perturbations in $G$ and $\rL$ and
when they are neglected. For $\nu<0$, our model predicts more growth
than the $\CC$CDM model. When $\delta G=0$, this is just due to the
fact that $\rho_\CC$ is decreasing when we go into the past, so its
repulsive effect diminishes. When considering the perturbations in
$G$ and $\Lambda$, the deviation with respect to the standard case
increases even more, as already commented in Fig.~\ref{fig2}a. The
opposite situation occurs for $\nu>0$; here the suppression of
growth is attributed to the enhanced repulsion of matter associated
to a positive $\rL$, which is increasing with $z$. Such suppression
is larger when we include the $G$ and $\CC$ perturbations. A
detailed study of these effects for model ii) of section
\ref{sec:GenericModels} was performed in Ref.\cite{Grande:2007wj},
in an effective approach with no perturbations in the CC.

\begin{figure}
\includegraphics[width=0.49\textwidth]{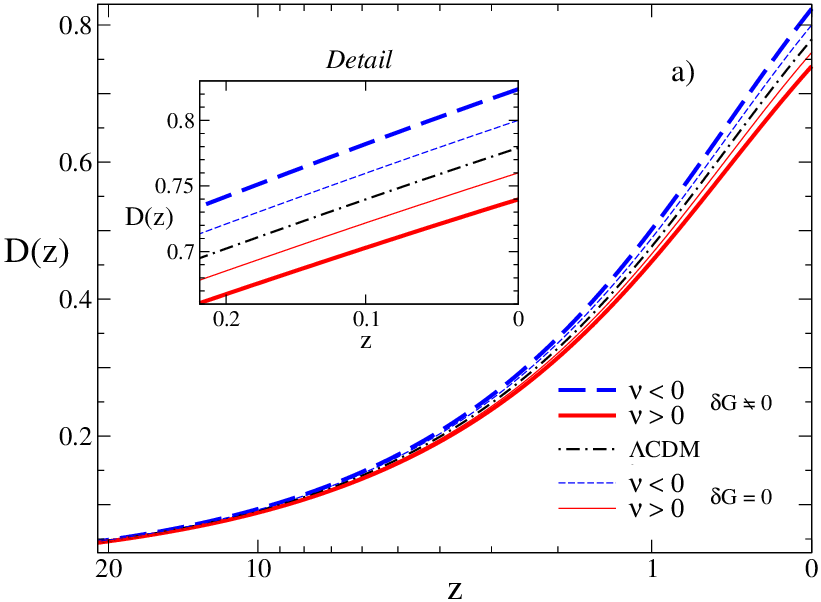}
\includegraphics[width=0.49\textwidth]{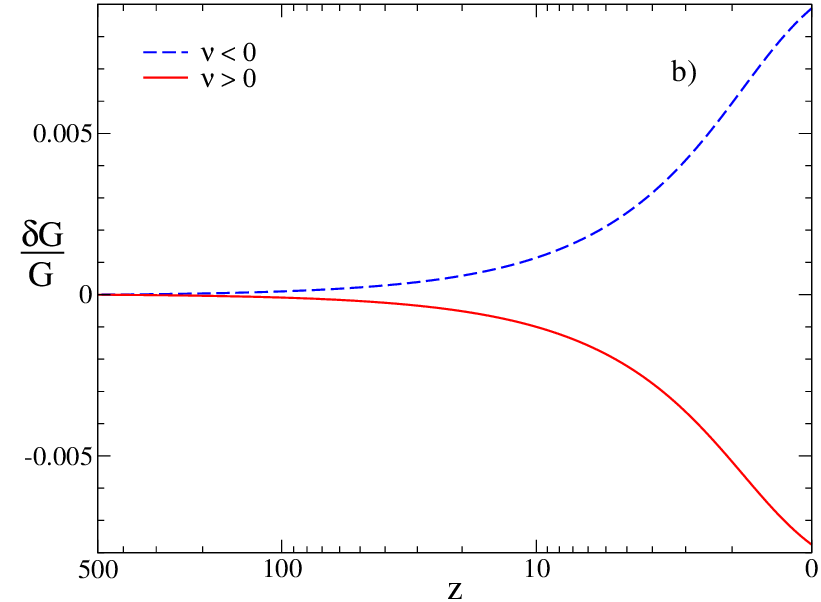}
\caption{Evolution of the perturbations for the QFT model with
running $\CC$ and $G$, using the same values of $\Omo$ and $\nu$ as
in Fig.~\ref{fig1}; a) The growth factor $D(z)$, showing an
enhancement (suppression) of the growth with respect to the $\CC$CDM
case for negative (positive) $\nu$. The effect is present even when
we neglect the perturbations in $G$ and $\rL$, but it is larger if
we consider them; b) Evolution of the perturbations in $G$, under
the natural assumption that they are negligible at $z_i=500$. The
perturbations remain unimportant until very recent times, when
$\tilde{\Om}_m$ begins to depart significantly from 1 (cf.
Fig.~\ref{fig1}). Remembering that $\delta_\CC=-\delta G/G$, we can
explain the enhancement (suppression) of matter perturbations
observed in a) on account of the opposite sign of
$\delta_\CC=\delta\rL/\rL$, see Eq.\,(\ref{diff1}).}\label{fig3}
\end{figure}

In Fig.~\ref{fig3}b, the evolution of $\delta G/G$, as obtained from
Eq.\,(\ref{diff2}), is depicted. This equation only determines
$\delta G'$, so we need to give an initial condition for $\delta G$
at $a=a_i$. In order to do so, we note that the fact that
$\tilde{\Om}_m(z)\simeq 1$ in the past (reflected in
Fig.~\ref{fig1}a) ensures that the perturbations in $G$ are not
playing any important role, according to our discussion at the end
of section~\ref{sec:general_perturb}. Therefore, the most natural
condition seems to be $\delta G(a_i)=0$, and, besides, we expect
$\delta G/G$ to remain negligible until very recent times
($z\lesssim10$), when $\tilde{\Om}_m$ begins to depart from 1. This
is indeed what can be seen in Fig.~\ref{fig3}b, giving additional
support to our assumption. Let us remind from (\ref{prime}) that
$\delta_\CC=-\delta G/G$. Thus, for $\nu<0$ we have $\delta G>0$ and
hence $\delta_\CC<0$, which explains the further enhancement in the
growth of matter perturbations with respect to the case with $\delta
G=\delta\rL=0$.

The main conclusion of this section is that for $\nu$ of order
$10^{-3}$ or smaller, our model is in perfect agreement with recent
data on the normalization of the power spectrum. For the time being,
there is still some controversy on the value of $\Omo$ resulting
from the power spectrum shape measured by the two main galaxy
surveys. For a given value of $\Omo$, the power spectrum in our
model presents, in very good approximation, the same shape as in the
$\CC$CDM model. However, a non-vanishing $\nu$ could become manifest
through a difference in the amount of growth. Therefore, a future
simultaneous precision cosmology measurement of the shape and
amplitude (normalization) of the matter power spectrum should have
the potential to discriminate between a model of variable $\rL$ and
$G$, and the $\CC$CDM.

In the following section, we show that $\nu\sim\mathcal{O}(10^{-3})$
is also the maximum allowed value that we would expect from
arguments related to primordial nucleosynthesis, fact that
reinforces the validity of this bound.

\subsection{Constraints from primordial nucleosynthesis}\label{sec:num_nuc}

Big Bang nucleosynthesis (BBN) can also provide limits on the
possible variation of Newton's coupling $G$. The BBN predictions for
the light element abundances are sensitive to a number of
parameters, such as the baryon-to-photon ratio $\eta=n_B/n_{\gamma}$
(in view of the fact that the nuclear reaction rates depend on the
nucleon density) or the value of the Hubble function at the
nucleosynthesis time, $H_N\equiv H(z=z_N)$ (given that the expansion
rate competes against the reaction rates). Since $\eta\propto\Om_B$
can be accurately determined through CMB measurements, one can use
the observed abundances to constrain the expansion rate, and since
$H\propto\sqrt{G}$ these constraints can be directly
translated\footnote{Provided that the total energy density at
$z=z_N$ is approximately the same as in the standard case. Eq.
(\ref{OLnegli}) shows that this condition is satisfied in our
model.} into bounds on $g_N\equiv g(z_N)$, where $g(z)$ was defined
in Eq.\,(\ref{Eg}).

The current constraints on $g_N$ available in the literature usually
make use either of the deuterium abundance ($D/H$) or the $^4$He
mass fraction ($Y_p$). Deuterium has the advantage that it is not
produced in significant quantities after nucleosynthesis; its
primordial abundance can be determined in a quite precise manner by
studying the spectrum of the light from distant quasars, which
exhibits an absorption line corresponding to the deuterium present
in (high-redshift) intervening neutral hydrogen systems. This turns
deuterium into an excellent probe of the universe at the time of
BBN. However, its predicted abundance depends much more strongly on
$\eta$ than on $H_N$. As a consequence, the constraints on $g_N$
based on deuterium measurements are not very stringent. For
instance, in \cite{Copi:2003xd} it is found $g_N =
1.01^{+0.20}_{-0.16}$ at the $68.3\%$ confidence level or:
\begin{equation}
g_N = 1.01^{+0.42}_{-0.30} \quad (95\%)\,.\label{bound1}
\end{equation}
The abundance of $^4$He is much more sensitive to the expansion
rate, but accurate observational values are difficult to obtain,
since there are many potential sources of systematic uncertainties
\cite{Olive:2004kq}. As a result, a wide range of values for $Y_p$
can be found in the literature, see e.g. Table 12 in
\cite{Iocco:2008va}. In this reference, the authors used
$Y_p=0.250\pm0.003$ (in fact, they use $Y_p$ in combination with
$D/H$, although the dominant effect is that of helium
\cite{Iocco:2008va}) to obtain (see section 8.2.5 of the latter):
\begin{equation}
0.964<g_N<1.086 \quad (95\%)\,,\label{bound2}
\end{equation}
whereas in \cite{Cyburt:2004yc} a more conservative range for the
$^4$He mass fraction is adopted ($Y_p=0.2495\pm0.0092$), leading to:
\begin{equation}
0.9<g_N<1.13 \quad (68\%)\,.\label{bound3}
\end{equation}
Applying (\ref{bound1})-(\ref{bound3}) to our model, we can
effectively constrain our parameter $\nu$. In order to do so, we
compute $g_N$ from (\ref{GH}):
\begin{equation}
g_N=\frac{1}{1+\nu\ln\left(H^2_N/H_0^2\right)}\simeq\frac{1}{1+\nu\ln\left[\OR^0(1+z_N)^4\right]}\,,
\end{equation}
where $\OR^0$ is the radiation energy density fraction at present,
we have neglected both the matter and CC contributions to the
expansion rate at $z=z_N$, and is evident that $g_N$ can be
neglected as well in the logarithm. Taking $\OR^0\sim10^{-4}$ (which
includes photons and three light neutrino species) together with
$z_N\sim10^9$, and comparing the last expression to
(\ref{bound1})-(\ref{bound3}), we find:
\begin{eqnarray}
\textrm{From (\ref{bound1})}\quad\longrightarrow\quad-4.1\times 10^{-3}&\lesssim\nu\lesssim&5.5\times  10^{-3}\nonumber\\
\textrm{From (\ref{bound2})}\quad\longrightarrow\quad-1.1\times 10^{-3}&\lesssim\nu\lesssim&5.1\times  10^{-4}\nonumber\\
\textrm{From (\ref{bound3})}\quad\longrightarrow\quad-1.6\times
10^{-3}&\lesssim\nu\lesssim&1.5\times  10^{-3}\nonumber
\end{eqnarray}
Let us remind that (\ref{bound2}) was derived from a value for $Y_p$
with a (possibly unrealistic) very small error and that
$(\ref{bound3})$ is a 68\% value (whereas the other two limits are
given at the 95\% confidence level.) In any case, the conclusion
arising from this nucleosynthesis analysis seems to be that the
parameter $\nu$ can be, at most, of order $10^{-3}$. This is in
complete agreement with the constraint on $\nu$ obtained in
section~\ref{sec:num_str} from structure formation considerations.
Therefore, we have arrived to the same result by using two very
different methods, which gives additional credit to our conclusions.


\section{Conclusions}

\qquad In this paper, we have derived the general set of
cosmological perturbation equations for FLRW models with variable
cosmological parameters $\rL$ and $G$ in which matter is covariantly
conserved. To our knowledge, this is the first time that a complete
set of coupled differential equations for $\delta\rL$ and $\delta G$
is presented in the literature. We have shown that the linear growth
of matter perturbations of this {model, $D(a)\propto\delta_m(a)$, is
independent} of the wave number $k$. Adding this property to the
fact that we generally expect these models to present the same
transfer function as the $\CC$CDM}, the scaling dependence and hence
the shape of the power spectrum will not change with the {late time
evolution} and it will coincide with that of the $\Lambda$CDM model.
This fact is remarkable as it is in contrast to the situation, more
frequently studied in the literature, in which the time-evolving
vacuum models exchange energy with matter at fixed Newton's coupling
$G$\,\cite{BPS09}.

We have exemplified the difference in the power spectrum of running
vacuum models, with and without matter conservation, by considering
the class of cosmological models characterized by the running CC law
(\ref{RGlaw}), which we call quantum field vacuum models because the
evolution of the CC in them is of the form that one would naturally
expect from QFT, and more specifically from the renormalization
group evolution of the cosmological parameters. Such quantum field
vacuum models depend on a single parameter $\nu$, which acts as the
$\beta$-function for the running of the vacuum energy density
$\rL=\rL(H)$ and the running gravitational coupling $G=G(H)$. It is
interesting to note that while the running of the vacuum energy
$\rL(H)$ is quadratical in the expansion rate, the running of $G(H)$
is logarithmic in $H$, Eq.\,(\ref{GH}), and therefore much milder.
For $\nu>0$, the running of $G$ is asymptotically free, hence $G$
decreases (resp. increases) logarithmically  with the redshift
(resp. with the expansion), whereas for $\nu<0$ it decreases with
the expansion. The background evolution of these models was studied
in \,\cite{SSS04,Fossil07}.

In the present paper, we have concentrated on the study of the
cosmic perturbation equations of the running $\rL(H)$ and $G(H)$
models. After confronting the predicted matter growth with the
various cosmological data, we have found that the final region of
the parameter space is a naturalness region in which $\nu$ could be,
in absolute value, of order $10^{-3}$ at most. We have also checked
that this bound is compatible with the restriction on $G$ variation
that follows from primordial nucleosynthesis. Remarkably enough, the
results emerging from the perturbations analysis are derived from
the amplitude of the power spectrum rather than from its shape.
Numerically, the obtained upper bound on $\nu$ is perfectly
compatible with the quantum field theoretical definition of this
parameter, see  Eq.\,(\ref{nu}), and it implies that the mass scales
that would enter the quantum running of the cosmological parameters
could be one order of magnitude below the Planck scale, at most. The
result $|\nu|<{\cal O}(10^{-3})$ is also compatible with previous
analyses, using various independent procedures, of models with
running $\rL$ in which the vacuum can decay into
matter\,\cite{BPS09,Fabris:2006gt,Grande:2007wj,vdecay}. This decay
feature, however, is impossible for the models with running $\rL(H)$
\textit{and} $G(H)$ that we have studied here, and this is the basic
reason why they can exhibit the same power spectrum profile as the
$\CC$CDM. We cannot exclude that this property could have been
responsible for the fact that we have observationally missed this
fundamental possibility up to now. As we have emphasized, we expect
that these models should eventually be testable in the next
generation of precision cosmology observations from the analysis of
the spectrum amplitude, rather than of the spectral shape.

\vskip 0.5cm

{\large\bf Acknowledgments.} JG and JS have been supported in part
by MEC and FEDER under project FPA2007-66665, by the Spanish
Consolider-Ingenio 2010 program CPAN CSD2007-00042 and by DIUE/CUR
Generalitat de Catalunya under project 2009SGR502. JF thanks CNPq
and FAPES, and ISh is grateful to CNPq, FAPEMIG, FAPES and ICTP for
the partial support.

\vspace{0.5cm}

\section{Appendix: cosmic perturbations in the Newtonian gauge}\label{sec:Newtonian}
In this appendix, we sketch the derivation of the perturbation
equations in the Newtonian or longitudinal
gauge\,\cite{Mukhanov:1990me}. Our aim is to explicitly check that
the same physical consequences that we have derived in section
\ref{sec:general_perturb} for the synchronous gauge are recovered in
another gauge. In this way, we confirm in our particular context the
general expectation that calculations of cosmic perturbations at
deep sub-horizon scales should not present significant gauge
dependence.

The most general perturbation of the spatially flat FLRW metric can
be conveniently written as follows \cite{Mukhanov:1990me}
\beq\label{FLRWpert} ds^2
=a^2(\eta)[(1+2\psi)d\eta^2-\omega_{i}\,d\eta\,dx^i -\,
\left((1-2\phi)\delta_{ij}+\chi_{ij}\right)dx^{i}dx^{j}]\,, \eeq
where $\eta$ is the conformal time, defined through $d\eta=dt/a$.
The above metric is expressed in the notation of \cite{LXCDM08} and
consists of the $10$ degrees of freedom associated to the two scalar
functions $\psi$, $\phi$, the three components of the vector
function $\omega_i\, (i=1,2,3)$, and the five components of the
traceless second-rank tensor $\chi_{ij}$. Clearly, the synchronous
gauge that we used in section \ref{sec:general_perturb} is obtained
by setting $\psi=0$, $\omega_i=0$ and absorbing the function $\phi$
into the trace of $\chi_{ij}$, which then contributes six degrees of
freedom. In this gauge, the metric part of the analysis of cosmic
perturbations is tracked by the nonvanishing trace of the metric
disturbance: $ h\equiv g^{\mu\nu}\,h_{\mu\nu}=
 g^{ij}\,h_{ij}=-{h_{ii}}/{a^2}$. The corresponding
equations have been presented in detail in section
\ref{sec:general_perturb} after defining the ``hat variable'':
$\hat{h}=-{\partial h}/{\partial t}=\partial
\left(h_{ii}/a^2\right)/\partial t$.

An alternative possibility is to use the (conformal) Newtonian
gauge\,\cite{Mukhanov:1990me,Ma:1995ey}. In this case, we set
$\omega_{i}=\chi_{ij}=0 \,(\forall i,j)$ in Eq.\,(\ref{FLRWpert}). A
useful feature of this gauge is that the metric is diagonal and
another is that, in the Newtonian limit, $\psi$ plays the role of
the gravitational potential. It is now straightforward to repeat the
calculation of the perturbation equations in a similar way as in
section \ref{sec:general_perturb}. In the absence of anisotropic
shear stress (which is always the case with non-relativistic
particles, such as baryons and dark matter), the perturbed
Einstein's equations for $i\neq j$ give
$\psi=\phi$\,\cite{Ma:1995ey}, see also \cite{Bean:2010zq}.
Indicating by a prime the derivatives with respect to the scale
factor ($f'\equiv df/da$), the final perturbation equations in that
gauge read as follows:
\begin{eqnarray}
&&k^2\phi=-4\pi a^2\left\{G\left[\delta\rho_m+\delta\rho_\Lambda+3H\rho_m\frac{a^2}{k^2}\theta\right]+(\rho_m+\rho_\Lambda)\delta G\right\}\,,\nonumber\\
&&\delta\rho'_m+\rho_m\left(\frac{\theta}{aH}-3\phi'\right)+\frac{3}{a}\delta\rho_m=0\,,\nonumber\\
&&\theta'+\frac{2}{a}\theta=\frac{k^2}{Ha^3}\phi\,,\label{final}\\
&& \delta G'(\rho_m+\rho_{\Lambda})+\delta G \rho'_{\Lambda}+ G'(\delta\rho_m+\delta\rho_{\Lambda})+ G\delta\rho'_{\Lambda}=0\,,\nonumber\\
&&G\delta\rho_{\Lambda}+\rho_{\Lambda}\delta
G+\left(\frac{Ha^3}{k^2}\right)\,\rho_m G'\theta= 0\,.\nonumber
\end{eqnarray}
Notice that the role of $\hat{h}$ in the synchronous gauge is taken
up here by the variable $\phi=\psi$, which turns out to satisfy an
algebraic rather than a differential equation. Worth noticing is
also the fact that the last two equations of (\ref{final}) are
identical to equations (\ref{fin4}) and (\ref{fin5}) respectively,
except that in the present case we used derivatives with respect to
the scale factor rather than with respect to the cosmic time (as in
this way length scales can be better compared with the horizon).

We are now ready to show that for deep sub-Hubble (sub-horizon)
perturbations, i.e. those satisfying $k/(aH)\gg1$, these equations
lead to the same second and third-order differential equations for
the normalized matter overdensity $\delta_m=\delta\rho_m/\rho_m$,
which we have previously found within the synchronous gauge in
section \ref{sec:general_perturb}. To start with, the first equation
of (\ref{final}) can be cast as:
\begin{equation}
\phi=-\frac{3}{2}\frac{H^2a^2}{k^2}\left\{\tilde{\Omega}_m\,\delta_m+\tilde{\Omega}_\Lambda\,{\delta_\Lambda}
+\,\frac{\delta G}{G}\right\}\,,\label{alge2}
\end{equation}
where $\delta_\CC=\delta\rho_\CC/\rho_\CC$, and of course we use the
same notation as in section \ref{sec:general_perturb} concerning the
parameters $\tilde{\Omega}_m$ and $\tilde{\Omega}_\Lambda$. In the
previous equation, we have dropped the term $3H
\rho_m\,{a^2}\,\theta/{k^2}$ since it is negligible for sub-Hubble
perturbations. In this same regime of perturbations,
Eq.(\ref{alge2}) obviously implies that $\phi\ll \delta_m$. Next,
let us note that the second equation of (\ref{final}) can be
simplified by using the conservation of matter, giving:
\begin{equation}
\delta'_m+\frac{\theta}{aH}=3\phi'\,.\label{neweqlong}
\end{equation}
Differentiating this equation and using the third one in (\ref{final}) to substitute for $\theta'$,
we find:
\begin{equation}
\delta_m''-\frac{\theta}{aH}\left(\frac{3}{a}+\frac{H'}{H}\right)=3\phi''-\left(\frac{k}{H
a}\right)^2\ \frac{\phi}{a^2}\,. \label{inter}
\end{equation}
Furthermore, if we eliminate $\theta$ from this equation by means of (\ref{neweqlong}) and also $\phi$ using
(\ref{alge2}), we get:
\begin{equation}
\delta_m''-3\phi''+(\delta_m'-3\phi')\left(\frac{3}{a}+\frac{H'}{H}\right)=\frac{3}{2a^2}\left\{\tilde{\Omega}_m\delta_m+\tilde{\Omega}_\Lambda\delta_\Lambda
+\frac{\delta G}{G}\right\}\,.\label{inter3}
\end{equation}
Again we can neglect the terms that are small at sub-Hubble scales;
for example, in the previous equation we can neglect $\phi'$ ($\phi''$) as
compared to $\delta_m'$ ($\delta_m''$) on account of the previously discussed
relation $\phi\ll \delta_m$. In this way, we arrive at:
\begin{equation}
\delta_m''+\left(\frac{3}{a}+\frac{H'}{H}\right)\delta_m'=\frac{3}{2a^2}\left\{\tilde{\Omega}_m\delta_m+\tilde{\Omega}_\Lambda\delta_\Lambda
+\frac{\delta G}{G}\right\}\,.\label{inter3b}
\end{equation}
The almost final step is to substitute $\delta_\Lambda$ by means of
the last equation of (\ref{final}). Once more, we can argue (as we
did in section \ref{sec:general_perturb}) that apart from the
damping factor $Ha/k\ll 1$ for the sub-Hubble modes, there are
additional suppression effects, such as the presence of $G'$ which
is small as compared to $G$, and hence $(G'/G)\theta$ can be
considered almost second-order. Therefore, the last equation of
(\ref{final}) boils down in practice to
$G\delta\rho_{\Lambda}+\rho_{\Lambda}\delta G=0$, which simply
renders $\delta_\Lambda=-\delta G/G$. Using also the sum rule
(\ref{tildeOmegas}), the relation (\ref{inter3b}) finally takes on
the simpler form
\begin{equation}\label{deltasecondorder}
\delta_m''+\left(\frac{3}{a}+\frac{H'}{H}\right)\delta_m'=\frac{3\tilde{\Omega}_m}{2a^2}\left(\delta_m+\frac{\delta
G}{G}\right)\,.
\end{equation}
This equation exactly coincides with Eq.(\ref{diff1}), showing
indeed that our result was not gauge-dependent. Obtaining the
third-order equation for $\delta_m$ is now straightforward. From the
last two equations of (\ref{final}) one can show that
\begin{equation}
\delta_m=-\frac{(\delta G)'}{G'}\,.
\end{equation}
From here we proceed as in section \ref{sec:general_perturb} for the
synchronous gauge, i.e. we differentiate Eq.(\ref{deltasecondorder})
with respect to the scale factor and we reach once more the third
order differential equation (\ref{supeq}). (q.e.d.)

Some discussion is now in order. Let us first note that the variable
cosmological constant scenario ii) of section
\ref{sec:GenericModels}, whose perturbation equations were
originally analyzed in \cite{Fabris:2006gt} in the synchronous
gauge, has been reanalyzed in the Newtonian gauge in
\cite{Toribio09} and the same conclusions have been attained. We
should also mention a very recent paper
\cite{Perivolaropoulos:2010dk} where a nice discussion is made on
comparing the synchronous and Newtonian gauges in the context of a
model with self-conserved DE and without DE perturbations. As it is
well-known, in that case the two gauges give the same evolution for
the cosmological perturbations for scales well below the horizon --
see also\,\cite{Ma:1995ey,Yoo:2009au}. This feature was expected in
our model too, even if it is a bit more complicated than the one
considered in\,\cite{Perivolaropoulos:2010dk} (our DE is also
self-conserved, but we do have both DE and $G$ perturbations), and
in fact the physical discussion about the gauge differences is very
similar. In the same reference, it is also pointed out that the
$k$-dependent effects are negligible for large values of $k$,
although they could start to be appreciable for $k<0.01 h$ (or,
equivalently, for length scales larger than $\ell\sim 1/k\simeq
100h^{-1}$ Mpc.).

The issues about gauge dependence are of course important in
general, because scale-dependent effects may change from gauge to
gauge. For instance, let us note that there is no counterpart
(within the synchronous gauge analysis considered in section
\ref{sec:general_perturb}) of the scale-dependent effects introduced
by the first equation (\ref{final}). At the same time, there may be
scale dependent features which are tied to the particular
cosmological model under consideration, and they need not show up in
another model, even if we compare them within the same gauge.
Indeed, take once more the model discussed in Ref.
\cite{Perivolaropoulos:2010dk}, where the DE is self-conserved and
there are no perturbations in this variable. In such situation, the
cosmological perturbations of matter in the synchronous gauge turn
out to be exactly scale-independent (without any approximation). In
our case, in contrast, there is some residual scale dependence in
the synchronous gauge, although it can be shown to be negligible --
as we have extensively discussed in section
\ref{sec:general_perturb}. Gauge differences, however, can be
significant in certain cases, specially for calculations involving
very large scales, showing that in general it is non-trivial to
relate the perturbation variables to observable quantities in
different gauges\,\cite{Ma:1995ey,Yoo:2009au}. In that case, one
must keep the appropriate $k$-dependence (as e.g. in the Newtonian
gauge\,\cite{Perivolaropoulos:2010dk}) or resort to a
gauge-invariant formalism\,\cite{Bardeen80,Kodama84}. However, for
the range of wavenumbers usually explored in the analysis of the
matter power spectrum (viz. $0.01h$ Mpc$^{-1} < k < 0.2h$
Mpc$^{-1}$\,\cite{Cole:2005sx}), the gauge differences, and in
particular the $k$-dependence, should be small enough so as to be
licitly neglected. In other words, for such scales, the sub-Hubble
approximation that we have taken in this Appendix holds good and the
calculations in the two gauges are found to be exactly coincident,
as we expected.

Finally, let us note that the reason why the physical discussion can
be more transparent in the Newtonian gauge is because the time
slicing in this gauge respects the isotropic expansion of the
background. The synchronous gauge, instead, corresponds to free
falling observers at all points, entailing that its predictions are
relevant only to length scales significantly smaller than the
horizon, but at these scales it renders the same physics as the
Newtonian
gauge\,\cite{Ma:1995ey,Perivolaropoulos:2010dk,Yoo:2009au}.


\end{document}